


\def\today{\ifcase\month\or
 January\or February\or March\or April\or May\or June\or
 July\or August\or September\or October\or November\or December\fi
 \space\number\day, \number\year}

\font\BF=cmbx10 scaled \magstep 3
 5
\let\1\sp
\def\bigstar{{\BF *}}

\let\draft1  

\font\eightrm=cmr8

\overfullrule=0pt
\hbadness 10000
\vbadness 10000
\ifx\magn1\else\magnification 1200 \fi
\baselineskip=12pt
 \hsize=16.5truecm \vsize=21 truecm \hoffset=.1truecm
\parskip=14 pt

\def\draftonly#1{\ifx\draft1{\rm \lbrack{\ttraggedright {#1}}
                    \rbrack\quad}                \else\fi}


\newcount\refno \refno=1
\def\eat#1{}
\def\labref #1 #2#3\par{\def#2{#1}}
\def\lstref #1 #2#3\par{\edef#2{\number\refno}
                              \advance\refno by1}
\def\txtref #1 #2#3\par{  \def#2{\number\refno
      \global\edef#2{\number\refno}\global\advance\refno by1}}
\def\doref  #1 #2#3\par{{\refno=0  \eat{#2} \if0#2\else
     \vbox {\item {{\bf\lbrack#2\rbrack}} {#3\par}\par}
     \vskip\parskip \fi}}
\let\REF\lstref    
\def\Jref#1 "#2"#3 @#4(#5)#6\par{#1:\
    #2,\ {\it#3}\ {\bf#4},\ #6\ (#5)\par}
\def\Bref#1 "#2"#3\par{#1:\ {\it #2},\ #3\par}
\def\Gref#1 "#2"#3\par{#1:\ ``#2'',\ #3\par}
\def\tref#1{{\bf\lbrack#1\rbrack}}

\def\idty{{\leavevmode{\rm 1\ifmmode\mkern -5.4mu\else
                                            \kern -.3em\fi I}}}
\def\Ibb #1{ {\rm I\ifmmode\mkern -3.6mu\else\kern -.2em\fi#1}}
\def\Ird{{\hbox{\kern2pt\vbox{\hrule height0pt depth.4pt width5.7pt
    \hbox{\kern-1pt\sevensy\char"36\kern2pt\char"36} \vskip-.2pt
    \hrule height.4pt depth0pt width6pt}}}}
\def\Irs{{\hbox{\kern2pt\vbox{\hrule height0pt depth.34pt width5pt
       \hbox{\kern-1pt\fivesy\char"36\kern1.6pt\char"36} \vskip -.1pt
       \hrule height .34 pt depth 0pt width 5.1 pt}}}}
\def\Ir{{\mathchoice{\Ird}{\Ird}{\Irs}{\Irs} }}
\def\ibb #1{\leavevmode\hbox{\kern.3em\vrule
     height 1.5ex depth -.1ex width .2pt\kern-.3em\rm#1}}
  \def\Rl {{\Ibb R}}

\def\lessblank{\parskip=5pt \abovedisplayskip=12pt
          \belowdisplayskip=12pt }
\outer\def\iproclaim #1. {\vskip0pt plus20pt \par\noindent
     {\bf #1.\ }\begingroup \interlinepenalty=250\lessblank\sl}
\def\eproclaim{\par\endgroup\vskip0pt plus30pt\noindent}
\def\proof#1{\par\noindent {\bf Proof #1}\          
         \begingroup
                     }
\def\QED {\hfill\endgroup\break
     \line{\hfill{\vrule height 1.8ex width 1.8ex }\quad}
      \vskip 0pt plus100pt}
\def\QEDD {\hfill\break
     \line{\hfill{\vrule height 1.8ex width 1.8ex }\quad}
      \vskip 0pt plus100pt}

\def\nl{\hfill\break}
\def\nll{\hfill\break\vskip 5pt}


\def\A1n{A_1\otimes\cdots\otimes A_n}

\def\Order{{\bf O}}

\def\bra#1{\langle #1\vert}

\def\imply{\Rightarrow}

\def\ket#1{\vert#1\rangle}
\def\ketbra#1#2{{\vert#1\rangle\langle#2\vert}}
\def\mod{{\rm mod}\,}

\def\Order{{\bf O}}


\def\supp{{\rm supp }}
\def\tover#1#2{{\textstyle{#1\over #2}}}

\def\Tr{\mathop{\rm Tr}\nolimits}

\def\phi{\varphi}            
\def\epsilon{\varepsilon}    
\def\el{l}

\def\A{{\cal A}} \def\B{{\cal B}}  
 \def\H{{\cal H}}  
\def\M{{\cal M}}   \def\Z{{\cal Z}}
\def\O{{\cal O}}

\def\E{{\Ibb E}}   

\def\d{{\rm d}}
\def\tprod{{\prod}^*}         
\def\bsig{\sigma\!\!\!\!\sigma}     
\def\sig{\vec{\bsig}}               
\def\Prob#1{\hbox{\rm Prob}(\hbox{\eightrm #1})}  
\def\Probin#1#2{\hbox{\rm Prob}_{#1}(\hbox{\eightrm #2})}
\def\Ind#1{\hbox{\rm \bf I}[\hbox{\eightrm #1}]}    
\def\ev{{\cal E}}
\def\event{\ev}
\def\model{spin $S$ model with interaction $-P^{(0)}$}
\def\models{spin $S$ models with interaction $-P^{(0)}$}
\def\Je{J_{\rm even}}
\def\Jo{J_{\rm odd}}

\def\sign{\hbox{sign}}
\def\even{$\langle\cdot\rangle_{\rm even}$}
\def\odd{$\langle\cdot\rangle_{\rm odd}$}



\def\ie{i.e.\ }               
\def\eg{e.g.\ }               


\def\om{\omega}



\def\es{{\bf S}}

\def\en{{\bf n}}

\def\x{{\bf x}}
\def\y{{\bf y}}
\def\bx{{\bf x}}
\def\by{{\bf y}}
\def\bz{{\bf z}}
\def\bu{{\bf u}}
\def\bv{{\bf v}}

\def\sh{{\hat \sigma}}
\def\bI{{\rm\bf I}}







\let\REF\lstref

\REF ACCN \ACCN \Jref
       M. Aizenman, J.T. Chayes, L. Chayes, and C.M. Newman
       "Discontinuity of the Magnetization in One-Dimensional
        $1/|x-y|^2$ Ising and Potts Models"
        J. Stat. Phys. @50(1988) 1--40

\REF AKN  \AKN \Gref
        M. Aizenman, A. Klein, and C. Newman
        "Percolation methods for disordered quantum Ising models"
        preprint

\REF  AL \AL \Jref
        M. Aizenman and E.H. Lieb
        "Magnetic Properties of Some Itinerant-Electron
        Systems at $T>0$"
        Phys. Rev. Lett. @65(1990) 1470--1473

\REF AM \AM \Jref
        M. Aizenman and Ph.A. Martin
        "Structure of Gibbs States of One-Dimensional
        Coulomb Systems"
        Commun. Math. Phys. @78(1980) 99--116

\REF AN \AN \Gref
         M. Aizenman and B. Nachtergaele
         "Geometric aspects of quantum spin systems II"
         in preparation

\REF Aff \Aff \Jref
         I. Affleck
        "Exact results on the dimerization transition in SU(n)
         antiferromagnetic chains"
         J. Phys. C: Condens. Matter @2(1990) 405--415

\REF AffL \AffL     \Jref
     I. Affleck and E.H. Lieb
     "A proof of part of Haldane's conjecture on quantum spin chains"
     Lett. Math. Phys. @12(1986) 57--69

\REF Baxa \Baxa \Bref
        R.J. Baxter
        "Exactly Solvable Models in Statistical Mechanics"
        Academic Press, London 1982

\REF Baxb \Baxb \Jref
         R.J. Baxter
         "Magnetisation discontinuity of the two-dimensional Potts
          model"
          J.~Phys. A: Math. Gen. @15(1982) 3329--3340

\REF BBa \BBa \Jref
     M.N. Barber and M.T. Batchelor
     "Spectrum of the biquadratic spin-1 antiferromagnetic chain"
      Phys. Rev. @B40(1989) 4621--4626

\REF BBb \BBb \Jref
      M.T. Batchelor and M. Barber
      "Spin-s quantum chains and Temperley-Lieb algebras"
      J.~Phys. A: Math. Gen. @23(1990) L15--L21

\REF BK \BK \Jref
        R.M. Burton and M. Keane
        "Density and uniqueness in percolation"
         Commun. Math. Phys. @121(1989 ) 501--505

\REF CNPR \CNPR \Jref
        A. Coniglio, C.R. Nappi, F. Peruggi, and L. Russo
        "Percolation and Phase Transitions in the Ising Model"
        Commun. Math. Phys. @51(1976)315--323

\REF CS \CS \Jref
     J.G. Conlon and J-Ph. Solovej
      "Random Walk Representations of the Heisenberg Model"
      J. Stat. Phys. @64(1991) 251--270

\REF  CF  \CF \Jref
       M.C. Cross and D.S. Fisher
       "A new theory of the spin-Peierls transition with special
        relevance to the experiments on TTFCuBDT"
       Phys. Rev. @B19(1979)402--419

\REF DLS  \DLS  \Jref
        F.J Dyson, E.H. Lieb, and B. Simon
        "Phase transitions in quantum spin systems with isotropic
         and non-isotropic interactions"
         J. Stat. Phys. @18(1978)335--383

\REF FNW \FNW \Jref
        M. Fannes, B. Nachtergaele , and R.F. Werner
        "Finitely correlated states for quantum spin chains"
         Commun. Math. Phys. @144(1992) 443--490

\REF For \For \Gref
        C.M. Fortuin:
        "On the random cluster model II and III"
        {\it Physica\/} {\bf 58}, 393--418 (1972) and
        {\it Physica\/} {\bf 59}, 545--570 (1972)

\REF FK \FKa \Jref
       C.M. Fortuin and P.W. Kasteleyn
       "On the random cluster model I"
        Physica @57(1972) 536--564

\REF FK \FKG \Jref
       C.M. Fortuin,  P.W. Kasteleyn, and L. Ginibre
       "Correlation inequalities on some partially ordered sets"
       Commun. Math Phys. @22(1971) 89--103

\def\FK{\FKa,\FKG,\For}

\REF Fur  \Fur \Bref
       H. Furstenberg
       "Recurrence in Ergodic Theory and Combinatorial Number
        Theory"
        Princeton University Press, Princeton 1981

\REF  GKR  \GKR \Jref
        A. Gandolfi, M. Keane,  and L. Russo
        "On the uniqueness of the infinite occupied cluster in dependent
        two-dimensional site percolation"
        Ann. Prob. @16(1988)1147--1157

\REF Geo  \Geo \Bref
        H.-O. Georgii
        "Gibbs measures and phase transitions"
        W. de Gruytter, Berlin - New York 1988

\REF Gin \Gin \Jref
        J. Ginibre
       "Reduced density matrices for the anisotropic Heisenberg model"
        Commun. Math. Phys.  @10(1968)140--154

\REF  GA  \GA  \Jref
      S.M. Girvin and D.P. Arovas
       "Hidden topological order in integer quantum spin chains"
       Physica Scripta @T27(1989)156--159

\REF GM \GM \Jref
       D. Goderis and C. Maes
       "Constructing quantum dissipations and their reversible
        states from classical interacting spin systems"
       Ann. Inst. H. Poincar\'e @55(1991) 805--828

\REF  HK  \HK  \Jref
      M. Hagiwara and K. Katsumata
      "Observation of $S=\tover12$ degrees of freedom in an undoped $S=1$
linear
      chain Heisenberg antiferromagnet"
      J. Phys. Soc. Japan. @61(1992)1481--1484

\REF  HKal2   \HKalb  \Jref
      M. Hagiwara, K. Katsumata, I. Affleck, B.I. Halperin, and J.P. Renard
      "Hyperfine structure due to the $S=\tover12$ degrees of freedom in
       an $S=1$ linear chain antiferromagnet"
      J. Mag. Mag. Materials @104-107(1992)839--840

\REF  HKal1  \HKala  \Jref
      M. Hagiwara, K. Katsumata, J.P. Renard, I. Affleck, and B.I. Halperin
      "Observation of $S=\tover12$ degrees of freedom in an $S=1$ linear
      chain Heisenberg antiferromagnet"
      Phys. Rev. Lett. @65(1990)3181--3184

\def\NENP{\HKala,\HK,\HKalb}

\REF Hal  \Hal  \Jref
     F.D.M. Haldane
     "Continuum dynamics of the 1-D Heisenberg antiferromagnet:
      identification with the O(3) nonlinear sigma model"
     Phys.Lett. @93A(1983)464--468

\REF HKW \HKW \Jref
       A. Hintermann, H. Kunz, and F.Y. Wu
        "Exact Results for the Potts Model in Two Dimensions"
        J. Stat. Phys. @19(1978) 623--632

\REF Hol \Hol \Jref
        R. Holley
        "Remarks on the FKG Inequalities"
        Commun. Math. Phys. @36(1974)277--231

\REF Ken \Ken \Jref
       T. Kennedy
      "Long range order in the anisotropic quantum ferromagnetic Hei\-sen\-berg
       model"
       Commun. Math. Phys. @100(1985) 447--462

\REF KL \KL \Jref
      T. Kennedy and E.H. Lieb
      "Proof of the Peierls Instability in One Dimension"
      Phys. Rev. Lett. @59(1987)1309--1312

\REF KLS \KLS \Jref
      T. Kennedy, E.H. Lieb, and B.S. Shastri
      "Existence of N\'eel order in some spin 1/2 Heisenberg antiferromagnets"
      J. Stat. Phys. @53(1988)1019--1030

\REF KT \KT \Jref
      T. Kennedy and H. Tasaki
      "Hidden Symmetry Breaking and the Haldane Phase in S=1 Quantum
       Spin Chains"
       Commun. Math. Phys. @147(1992)431--484

\REF KiTh \KiTh \Jref
       J.R. Kirkwood and L.E. Thomas
       "Expansions and Phase Transitions for the Ground State of Quantum
        Ising Lattice Systems"
       Commun. Math. Phys. @88 (1983) 569--580

\REF {Kl\"ua} \Klua \Jref
          A. Kl\"umper
         "New Results for q-state Vertex Models and
          the Pure Biquadratic Spin-1 Hamiltonian"
          Europhys. Lett. @9(1989) 815--820

\REF {Kl\"ub} \Klub \Jref
          A. Kl\"umper
         "The spectra of q-state vertex models and related
          antiferromagnetic quantum spin chains"
          J.Phys. A: Math. Gen. @23(1990) 809--823

\def\Klu{\Klua,\Klub}

\REF  KLMR \KLMR \Jref
         R. Koteck\'y, L. Laanait, A. Messager, and J. Ruiz
         "The q-state Potts Model in the Standard Pirogov-Sinai Theory:
          Surface Tensions and Wilson Loops"
          J.~Stat.~Phys. @58(1990) 199--248

\REF  KS \KS \Jref
         R. Koteck\'y and S.B. Shlosman
         "First-Order Phase Transitions in Large Entropy Lattice Models"
          Commun. Math. Phys. @83(1982) 493--515

\REF LM \LM \Jref
        E.H. Lieb and D.C. Mattis
        "Ordering Energy Levels of Interacting Spin Systems"
        J.~Math.~Phys. @3(1962) 749--751

\REF LSM \LSM    \Jref
   E. Lieb, T. Schulz, and D. Mattis
   "Two soluble models of an antiferromagnetic chain"
   Ann.Phys.(NY) @16(1961)407--466

\REF Mat1 \Mata  \Jref
      T. Matsui
      "Uniqueness of the Translationally Invariant Ground State
      in Quantum Spin Systems"
      Commun. Math. Phys. @126(1990) 453--467

\REF Mat2 \Matb  \Jref
     T. Matsui
     "On Ground State Degeneracy of $\Ir_2$-Symmetric Quantum
      Spin Models"
      Pub. RIMS (Kyoto) @27(1991)657--679

\REF New \New \Jref
         C.M. Newman
          "A general central limit theorem for FKG systems"
         Commun. Math. Phys. @91(1983)75--80

\REF dNR \DNR \Jref
      M. den Nijs and K. Rommelse
      "Preroughening transitions in crystal surfaces and valence bond phases
      in quantum spin systems"
      Phys. Rev. @B40(1989) 4709--4734

\REF PS \PS \Jref
     L. Pitaevskii and S. Stringari
     "Uncertainty principle, quantum fluctuations and broken symmetries"
     J. Low Temp. Phys. @85(1991)377--388

\REF Sha  \Sha \Jref
      B.S. Shastri
     "Bounds for correlation functions of the Heisenberg antiferromagnet"
     J.~Phys. A: Math. Gen. @25(1992)L249--253

\REF Sut \Sut \Jref
     B. Sutherland
     "Model for a multicomponent quantum system"
     Phys. Rev. @B12(1975)3795--3805

\REF Suto \Suto \Gref
     A. S\"ut\"o
    "Percolation transition in the Bose gas"
     to appear in J. Phys. A: Math.~Gen.

\REF Tho \Tho \Jref
     L. E. Thomas
     "Quantum Heisenberg ferromagnets and stochastic exclusion
      processes"
     J. Math. Phys. @21(1980) 1921--1924

\REF Toth \Toth \Gref
        B. T\'oth
       "Improved lower bound on the thermodynamic pressure of the spin 1/2
        Heisenberg ferromagnet"
       to appear in Lett. Math. Phys.

\REF Wu \Wu \Jref
        F.Y. Wu
        "The Potts model"
        Rev. Mod. Phys. @54(1982) 235--268


\vsize=22 truecm

\hrule height 0pt
\vskip 30pt plus30pt

\centerline{\BF Geometric Aspects of Quantum Spin States\bigstar}

\vskip 25pt plus25pt

\centerline{
Michael Aizenman$^{**}$\qquad and\qquad
Bruno Nachtergaele}
\centerline{
\vtop{\hsize = 6 truecm
\centerline{Department of Physics}
\centerline{ Princeton University}
\centerline{Jadwin Hall, P.O.Box 708}
\centerline{Princeton NJ 08544-0708, USA}
}}
\vskip 50pt plus50pt

\centerline{\bf Abstract}\nl
\centerline{\vtop{\hsize = 14 truecm \noindent
 A number of interesting features of the ground
states of quantum spin chains are analized with the help of a functional
integral
representation of the system's equilibrium states.  Methods of general
applicability
are introduced in the context of the SU($2S+1$)-invariant quantum spin-$S$
chains with the interaction $-P^{(0)}$, where $P^{(0)}$ is the projection onto
the singlet state of a pair of nearest neighbor spins.   The phenomena
discussed
here include: the absence of N\'eel order, the possibility of dimerization,
conditions for the existence of a spectral gap, and a dichotomy
analogous to one found by  Affleck and Lieb, stating that the systems exhibit
either slow decay of correlations or translation symmetry breaking.
 Our representation  elucidates the relation,  evidence for which
was found earlier, of the $-P^{(0)}$ spin-$S$ systems with the Potts and the
Fortuin-Kasteleyn random-cluster  models in one more dimension.
The method reveals the geometric aspects of the listed phenomena,
and gives a precise sense to a picture of  the  ground state in which
the spins are grouped into random clusters of zero total spin.   E.g., within
such structure  the dichotomy is implied by a topological argument,
 and the alternatives correspond to whether, or not, the clusters are of finite
mean length.
}}

\vfill

\line{\hfill June 14, 1993; revised September 24, 1993\qquad\quad}

\vfootnote*{Work supported in part by NSF Grant PHY-9214654.}
\vfootnote{**}{Also in the Mathematics Department.}
\vfootnote{}{E-mail: {\tt aizenman@phoenix.princeton.edu}}
\vfootnote{}{ \ \ \ \ \ \ \ \ \ \  {\tt bxn@math.princeton.edu}}

\vfill\eject

\def\tocs#1#2{\vskip .3cm%
   \line{#1 \dotfill \hbox{#2}}}%
\def\tocss#1#2{\vskip .3 cm%
   \line{\quad #1 \dotfill #2}}%
\def\tocsss#1#2{\vskip .3cm%
   \line{\quad\quad #1 \dotfill #2}}%
\phantom{ }

\vfill

\noindent
{\bf Table of Contents}

\vskip .7cm

\tocs{1. Introduction}{3}
\tocs{2. Quasi-state decompositions for equilibrium states of quantum spin
systems}{7}
\tocss{2.1. The functional integral representation for the spin-$S$ model
with interaction $-P^{(0)}$}{7}
\tocss{2.2. Poisson process representation of $e^{-\beta H}$}{11}
\tocss{2.3 The Q-S decomposition}{12}
\tocss{2.4. The spin-1/2 Heisenberg ferro- and antiferromagnet}{13}
\tocsss{2.4.a. The ferromagnet ($H^{F}$)}{14}
\tocsss{2.4.b. The antiferromagnet ($H^{AF}$)}{16}
\tocss{2.5. Structure of the quasi-states}{18}
\tocss{2.6. The SU(2s+1)-invariant spin-$S$ models with the interaction
$-P^{(0)}$}{19}
\tocs{3. Equivalence with to the 2-dimensional $(2S+1)^2$-state
Potts-models}{21}
\tocs{4. Finite systems and the thermodynamic limit}{29}
\tocs{5. Absence of N\'eel order}{33}
\tocs{6.  Dimerization versus Power Law Decay: a dichotomy}{35}
\tocss{6.1. The dichotomy}{36}
\tocss{6.2. The dimerization order parameter}{39}
\tocs{7. Decay of correlations in the \model}{42}
\tocs{Appendix I: Quasi-state decomposition for quantum states}{55}
\tocs{Appendix II: FKG structure and the Rising Tide Lemma}{58}
\tocs{References}{62}

\vfill\eject

\beginsection{1. Introduction}

There is a geometric aspect to the structure of the spin-spin
correlations found at low temperatures in a number of quantum
spin systems.  Our purpose is to introduce some generally
applicable tools for the analysis of such phenomena.  That is
done in the context of the SU($2S+1$) invariant models introduced
by Affleck and also studied by Batchelor and Barber, and Kl\"umper
\tref{\Aff,\BBa,\BBb,\Klu},
which include the
spin 1/2 Heisenberg antiferromagnet as a special case.

The systems considered here are one dimensional chains of spin $S$
variables, with the Hamiltonian
$$
H = -(2S+1)\sum_x J_x P^{(0)}_{x,x+1}
\eqno(1.1)$$
where $P^{(0)}_{x,y}$ is the orthogonal projection onto the singlet state of
two quantum spins of magnitude $S$, and $J_x>0$. The models with
translation invariant (all $J_x=J$) or staggered coupling constants
(possibly with two different values for $x$ even and odd) are of special
interest and some of our results are specific for these cases.
We will refer to the Hamiltonians (1.1) as the \models. The
explicit form of the interaction in the basis of eigenstates of $S^3$ is:
$$
(2S+1)P^{(0)}_{x,y}
=\sum_{\alpha,\beta=-S}^S
(-1)^{\alpha-\beta}\ketbra{\beta,-\beta}{\alpha,-\alpha}
\eqno(1.2)$$
The interaction can of course also be expressed as a polynomial in the
Heisenberg interaction $\es_x\cdot\es_{x+1}$. For $S=1/2$ and $S=1$ one obtains
$$
P^{(0)}_{x,y}=\cases{\tover14 -  \es_x\cdot\es_{x+1}  & for $S=1/2$\cr
\tover13(\es_x\cdot\es_{x+1})^2  -\tover13  & for $S=1 $\cr}
\eqno(1.3)$$
The analogous expressions for  general $S$ can be found, \eg , in \tref\Klub .

The phenomena which we shall address are:

{\it 1) The nature of the order parameters\/} which characterize the
possible occurence of symmetry breaking in the ground state.

{\it i.  N\'eel order.}  In higher dimensions such models may exhibit
N\'eel order in the ground state.  For the standard Heisenberg antiferromagnet
this
has been proved for dimensions $d\geq 3$ and also for $d=2$ if
$S\geq 1$ \tref{\DLS,\KLS}.
The representation introduced here permits to rule out
that possibility for the translation invariant models with interaction (1.1)
in one dimension, on the basis of known
results in percolation theory (in two dimensions).
(In one dimension, the representation relates N\'eel order
to a transient behavior in a system of random loops which form
the boundaries of the connected clusters of a random cluster model).

{\it ii.  Dimerization.}
The one dimensional models may, nevertheless, exhibit a
two-fold translation symmetry breaking, caused by dimerization.
While the interaction favors the pairing of neighboring
spins into singlet states, not all neighboring spins
can be paired simultaneously.
There are, of course, states - corresponding  to different
dimerizations of the lattice, in which half (or, on the lattice
$\Ir^d$, a fraction $1/(2d)$ ) of the
interaction terms are minimized.
While these are not true ground states, it turns out that
in one dimension for $S$ large enough ($S\geq1$)
this structure is present in the ground state, which
decomposes into a superposition of two partially dimerized states.
Spins on even sites have stronger correlations with their neighbors to
the right in one of the states, and to the left in the other.

The classical dimerization picture is too naive in two aspects:
1) the model's correlation functions extend beyond
nearest neighbors, and 2)  the  spins correlate in larger clusters than
pairs.  A virtue of the method employed here is that it permits
to describe this phenomenon (and the picture of the
correlated clusters) in explicit and precise terms.
In particular, we find the following behavior: in the state
where the spins on the even sites are more correlated with their neighbors
to the right one finds that with probability 1 some spins on the left
half-infinite chain $(-\infty,x]$ form a singlet with some spins on the
right half-infinite chain $[x+1,+\infty)$, for each $x$ even. In the same
state this probability is $<1$ for $x$ odd.

{\it 2) The ``dimerization versus power law decay'' dichotomy.}

The dimerization does not always occur.
However, we show that there is a dichotomy: the ground state
either dimerizes, or exhibits slow decay of correlations
(with $\sum_x\vert x \langle S^3_0 S^3_x\rangle\vert=+\infty$).
The dichotomy has the following geometric content.
When the (even) clusters of correlated spins are  tightly bound,
with only a finite number of clusters having the origin in their
span, then a topological argument implies that the
translation symmetry  has to be  broken.  The alternative is
that the clusters of correlated spins are loosely bound, with
the origin (as well as any other lattice site) belonging to
the span of infinitely many correlated pairs.  In that case,
the above sum of the correlation function diverges.
In fact, in our representation, that sum measures the number of
correlated spin pairs with $x<0$ and $y>0$.

The dichotomy discussed here is reminiscent of the one found
by Affleck and Lieb for the Heisenberg antiferromagnetic spin
chain with half-integer spins \tref\AffL .  However, unlike the
dichotomy of \tref\AL, the one discussed here is not restricted
to half-integer spin. The string order parameter
mentioned above is also a variant of one which has been
found relevant before in the context of the Heisenberg model.
In fact, the method introduced here applies to quite a broad
class of antiferromagnetic spin models, to be discussed
in a subsequent paper  \tref\AN .

{\it 3)  Relation with the Potts models.}

The ground states of
the 1D spin chains with the Hamiltonian (1.1), as well
as the Gibbs states $\exp(-\beta H)$ are related
to Potts model, which in case of the translation invariant
interaction are always at the self-dual point.
The existence of a relation was noted, at
the level of a similarity of the spectra of the relevant transfer
matrices, by Baxter \tref{\Baxa} for the spin 1/2 model,
and by Affleck (who introduced the general \model ),
Batchelor and Barber, and Kl\"umper \tref{\Aff,\BBb,\Klub}
for general spins.
The representation employed here makes this relation
very explicit.  In particular, the dimerization corresponds
to the existence
of long range order in the corresponding Potts model, and
the expectation values of any observable of the spin chain
can be expressed
in terms of quantities calculated within the Potts model.
The relation presented here extends also to models with
inhomogeneous couplings (for which the corresponding
Potts models are
no longer at their transition point), and thus extends
beyond the exactly soluble cases discussed in
\tref{\Baxa,\Aff,\BBb,\Klub}.

{\it 4)  Decay rate. }

Using the geometric representation, and the
FKG inequalities which it allows to bring to bear on the
problem, we derive an effective bound on the decay of
correlations of general observables in terms of the truncated two-point
function $\tau(x,y)$ of an associated two-dimensional $(2S+1)^2$-state
Potts model:
$$
\vert\langle A B_\bz\rangle\vert\leq
C_A C_B\sum_{x\in \overline\supp A
\atop y\in \overline\supp B_\bz}\tau(x,y)
\eqno(1.4)$$
For a more complete statement and the
notation see Theorem 7.2. Assuming that the truncated two-point function
of the two-dimensional Potts model in a magnetically ordered pure phase
always has exponential decay, this result implies the existence of a spectral
gap in the ground states of the \models\ whenever dimerization occurs.
The case of staggered couplings, $\{\Je,\Jo\}$,
is of  interest for the
discussion of the spin Peierls instability.  Our method
confirms the results obtained by Cross and Fisher
for the exponents describing the leading behavior of the
energy and the mass gap as a function
of $\vert \Je-\Jo\vert$ \tref\CF .

This paper serves as an introduction to a technique of
wider application, which is based on a decomposition of the Gibbs states
of a large class of quantum spin Hamiltonians as superpositions of
 what we call quasi-states (see Appendix I).
In a subsequent paper we show that such a representation exists for
any isotropic nearest neighbour interaction under the condition that
there are no frustration effects. No frustration essentially means that
the lattice is bipartite and the Hamiltonian has ferromagnetic interactions
only  within each
sublattice, and all interactions between spins of
different sublattices are antiferromagnetic. In case
there are only ferromagnetic interactions the bipartite structure is
irrelevant.

For the more general case, the method used here permits to
 give a natural
description of the occurrence of fractional spins at the edges of
finite chains. Let us just mention some results for the  spin-1
antiferromagnetic chain with Hamiltonian:
$$
H=\sum_x a \es_x\cdot\es_{x+1} + b(\es_x\cdot\es_{x+1})^2
\eqno(1.5)$$
with the coupling constants $a$ and $b$ satisfying $a\geq 0$ and
$b\leq 0$.
Again there is a dichotomy \tref\AN : under the assumption
of sufficiently fast decay of correlations (expected to be violated only
when  $a=-b$) either

\noindent
1) the ground state of the infinite system dimerizes, and thus
breaks the translation invariance of the
Hamiltonian,

\noindent
or,  2) finite pieces of the chain behave as if near each edge there
was an excess spin $S= \tover12$.

The latter case corresponds to the Haldane phase \tref\Hal\ and the spin-1/2 's
at the edges have been observed in electron spin resonance
experiments on NENP \tref{\NENP}.

\beginsection 2.  Quasi-state decompositions for equilibrium states of quantum
spin systems

In this section we derive the path integral representation for
the ground state, and  the equilibrium states, which is employed
in the derivation of the results described in the
introduction.  The discussion applies to  a more general class
of systems than those covered by equation (1.1).  We also
introduce here the notion of a `quasi - state decomposition'
of a quantum state.  Some of its basic properties are
presented in Appendix I.

Before turning to the derivation, let us state the net result
for the \models.

\noindent
{\bf 2.1 The functional integral representation for
the \models }\nl

Absorbing a convenient constant in its definition, the Hamiltonian
is now given by:
$$
H=\sum_x \{1-(2S+1)P^{(0)}_{x,x+1}\}
\eqno(2.1)$$
and we are considering finite chains of spins
of magnitude S. (The validity of Proposition 2.1 below is actually not
restricted to the one-dimensional case, though it requires the
model to be frustration free, i.e., to have a bipartite structure).

We denote by
$\sigma = \{\sigma_x\}$ a configuration of joint values for the
commuting family of observables $\{S^{(3)}_x\}$.
These configurations form a natural parametrization for an
orthonormal basis of the Hilbert space of the system's state
vectors.  We obtain the following representation for the matrix
elements, in this basis, of  the operator $e^{-\beta H}$.

For each specified pair of configurations, $e^{-\beta H}(\sigma',\sigma)$
is given by an integral over various possible histories
 of a time dependent
configuration $\sigma(t)=\{\sigma_x(t)\}$, with the time $t$ ranging over
the interval $[0,\beta]$, and
 $\sigma(0)=\sigma$ and  $\sigma(\beta)=\sigma'$.
What remains to
be specified is the description of the configurations
contributing to this integral, and the measure with which they
are integrated over.

The contributing spin configurations are piecewise constant in time.
When a change occurs,  the spins change simultaneously at a pair
of neighboring sites, but the two spins
are constrained to add to zero both before and after the change.
A useful description of the process is obtained by associating
with each spin configuration a collection of
`horizontal' bonds (in the space-time diagram in which time is in the
vertical direction)  linking the pairs of related sites in all the
discontinuity events.
For a technical reason, we find it convenient
to somewhat modify this relation, and consider the bonds
as enabling, rather than forcing the spin flips.  (This extension
yields a higher degree of independence in the measure seen below).
The integral over the time dependent spin configurations
is based on an integral over those
time indexed bonds, which we denote by  the symbol $\omega$.

We shall use the following symbol for the consistency
indicator function
$$
\Ind{$\sigma(\cdot)|\omega$} =\cases{1 & if
 ${\sigma_x(t)}$ satisfies the above described constraints
\cr
 & and all its discontinuities occur at bonds in $\omega$\cr
0& otherwise \cr}
\eqno(2.2)$$
and $\bI_{\rm per}[\sigma(\cdot)|\omega]$ which equals one if, in addition,
$\sigma(\cdot)$ is periodic in time ($\sigma_x(\beta)=\sigma_x(0)$).

The collection of all the spin configurations which are
consistent with a given bond
configuration $\omega$ is conveniently described with the aid of a
loop decomposition of the space-time diagram
(which forms a finite-volume subset of $\Ir\times [0,\beta]$ as
illustrated in Figure 1).   The loop to which a point $(x,t)$
belongs is found by moving from it along the vertical line
till a bond is reached. Upon reaching a bond, the path traverses
it, and then continues in the reversed time direction along the
vertical line to which it just crossed.
This procedure is continued until the path either closes
(by returning to its starting point), or reaches the time
$t=0$ or $t=\beta$.  For the time-periodic constraint, the loops
do not stop at $t=0,\beta$ but reemerge at the other end.
Following these instructions, space-time
is decomposed into a collection of lines which may form
either closed loops  or
lines connecting pairs of `boundary
sites' (at $t=0, \beta$) (in the non-periodic case).
Drawing each bond in duplicate,
the corresponding lines may be drawn so that they do not cross.
For a specified $\omega$, the consistent spin configurations
are
completely characterized by the condition that the staggered spins,
$(-1)^x \sigma_x(t)$ are constant
along each of the loops of $\omega$.
In particular, for the periodic constraint,
the number of consistent spin configurations is,
$(2S+1)^{\el_{\rm per}(\om)}$, with $\el_{\rm per}(\om)$ being the number of
loops.

The relevant measure for the time dependent spin configurations
can be constructed by means of a product measure, obtained by
the integration over $\omega$ with an auxilliary
Poisson process distribution, $\rho_{[0,\beta]}(d\omega)$,
and the discrete summation over the $(2S+1)^{\el_{\rm per}(\om)}$
consistent spin configurations (i.e., those with
$\Ind{$\sigma(.)|\omega$} = 1$).  The Poisson measure is
characterized by the condition that the mean bond density is
1, and that they occur
independently in disjoint regions.  The contribution of a given
bond configuration to the integral is enhanced by the factor
$(2S+1)^{\el_{\rm per}(\om)}$, and therefore its  effective
weight in the partition sum is given by the probability measure
$$
\mu_{\beta}(\d\om)=(\Z_\beta)^{-
1} \rho_{[0,\beta]}(\d\om)(2S+1)^{\el_{\rm per}(\om)}
\eqno(2.3)$$

\vfill\eject

\phantom{ }

\vskip 0pt plus 1filll

\noindent
{\bf Figure 1:} {\it A space-time configuration $\om$ for the $-P^{(0)}$
quantum spin chain,  at an inverse temperature $\beta$.}
As in a more general case, the spins
are correlated  within loops drawn by following the lines in the space-time.
A special feature of this interaction is that the loops can be viewed
as the boundaries of the  connected clusters of two
 random cluster models, dual to each other.
 The shaded areas are the connected $A$-clusters,
the connected $B$-clusters are left blank.   $A_1,A_2,B_1,B_2$
are the four independent regions surrounding a bond that appear
in the proof of the Euler relation.
The trace of the
loops on the $t=0$ line shows a decomposition of the spins into random
clusters of zero spin.

\vfill\eject

The situation is summarized in the following proposition.

\iproclaim Proposition 2.1.
For a  finite system with the Hamiltonian
(2.1):
\nl i)
$$  \langle \sigma' \mid e^{-\beta H} \mid \sigma \rangle =
\int \rho_{[0,\beta]}(\d\om)
\sum_{\sigma(\cdot) : \bI[\sigma(\cdot)\mid\omega]=1}
\bI[ \sigma(\beta)=\sigma',
\sigma(0)=\sigma]
\eqno(2.4)$$
ii) the partition function is given by:
$$
\Z_\beta=\Tr e^{\beta\sum_{x} ((2S+1)P^{(0)}_{x,x+1}-1)}=
\int\rho_{[0,\beta]}(\d\om) \, (2S+1)^{\el_{\rm per}(\om)}
\eqno(2.5)$$
iii) the equilibrium expectation values of observables
which are functions of
the operators ${S^{(3)}_{x}}$
can be expressed as
$$
{\Tr f(\{S_x^3\})e^{-\beta H}\over
\Z_\beta}
=\int\mu(\d\om)\, E_\om(f)
\eqno(2.6)$$
where $\mu(\d\om)=\Z_\beta^{-1} \rho_{[0,\beta]}(\d\om) \,
(2S+1)^{\el_{\rm per}(\om)}$ and the expectation functional $E_\om(f)$ is
obtained
by averaging, with equal weights, over all the spin
configurations consistent with $\omega$:
$$
E_\omega(f)={1\over (2S+1)^{\el_{\rm per}(\omega)}}
\sum_{\sigma : \bI[\sigma(\cdot)\mid\omega]=1} f(\sigma(t=0))
\eqno(2.7)$$
\eproclaim

In the above proposition and in the rest of this paper by
$\Ind{$\cdots$}$ we denote the indicator function of the event
described between the brackets.
Of course,  the objects $\Z_\beta,
\rho_{[0,\beta]}(\d\omega),E_\omega(f)$ and $\mu(\d\omega)$,
depend on the size of
the finite system and the magnitude $S$ of the spins.

The functionals $E_\omega$ can be extended to the full
algebra of observables (see Section 2.5 for explicit expressions).
Thus equation (2.6) is akin to a reperesentation of the equilibrium
state as a superposition of states.  That, however, is only partially
true.  The linear functionals do not have the full positivity
properties of quantum states.  Nevertheless, this point of view is
very useful, and is well justified in so far as the expectation values
of the special (but important) subalgebra of observables
$\{f(\{S_x^3\})\}$ is concerned.  We refer to such functionals as
quasi-states.  The notion is ellucidated in Appendix I.

As given by equation (2.7), in each quasi-state $E_\omega$
the joint distribution of the spins takes a very simple form.
In particular:
$$
E_\omega(S^3_x S^3_y)=(-1)^{\vert x-y\vert}
C(S)
\Ind{\hbox{
$(x,0)$ and $(y,0)$ are on the same loop} }
\eqno(2.8)
$$
with $C(S)={1\over 2S+1} \sum_{m=-S}^S m^2=\tover13 S(S+1)$.

Hence, the spin-spin correlation is proportional to the probability,
with respect to the effective probability measure on the space of
bond configurations, that
two sites are on the same loop of $\omega$:
$$
\langle S^3_x S^3_y \rangle = (-1)^{\vert x-y\vert}
C(S)   \Probin{\mu}{
$(x,0)$ and $(y,0)$ are on the same loop}
\eqno(2.9)$$

The rest of this section is devoted to the derivation
of Proposition 2.1, and of similar results for other systems
(e.g. the Heisenberg ferromagnet).
The discussion of the specific properties of the model with
$H=-\sum_x P^{(0)}_{x,x+1}$ is resumed in Section 3.

\noindent
{\bf 2.2 Poisson process representation of $e^{-\beta H}$}\nl

We now turn to the derivation of the functional integral
representation, which is done in the broader context of
operators of the form
$$
H_\Gamma=-\sum_{b\in\B} J_b h_b
\eqno(2.10)$$
where   $\Gamma$ is a (finite) collection of sites,
$\B$ is a  collection of
subsets of $\Gamma$, and for each $b\in\B$, $h_b$ is a self-adjoint
operator acting in the Hilbert space
 $\bigotimes_{i\in b} \H_i$, with $\H_i$ the state space at
the site  $i$,
and  $J_b$ are {\it non-negative \/}
coupling constants.
We refer to the sets $b\in\B$ as bonds, although for the
moment they are not required to be pairs of sites.

Thermal equilibrium states of the system, and its ground
state (approached in the limit $\beta\to\infty$), are
associated with the operator $e^{-\beta H}$.
Following is a general expansion of such
operators by means of integrals over a Poisson process.
The symbols $\rho$ and $\omega$ appearing here are defined afresh,
but their usage is consistent with the example discussed in
the previous subsection.
$$\eqalignno{
 e^{-\beta H}
&= e^{\beta \sum_{b\in\B} J_b}
\lim_{\Delta t\to 0} \left({\prod}_{b\in\B} e^{(-J_b+J_b h_b)
\Delta t}
\right)^{\beta/\Delta t}\cr
&= e^{\beta \sum_{b\in\B} J_b}\lim_{\Delta t\to 0}
\left({\prod}_{b\in\B} \{(1-J_b\Delta t)
+J_b  h_b\Delta t\}\right)^{\beta/\Delta t}\cr
&=e^{\beta \sum_{b\in\B} J_b}\int\rho^J_{[0,\beta]}(d\om)
\tprod_{(b,t)\in \om}
h_b&(2.11)\cr
}$$
where  $\om=\{(b_i,t_i)\}\subset \B\times[0,\beta]$
is a configuration of time indexed bonds, $\tprod$ is the
time ordered product:
$$
\tprod_{b\in\om}h_{b}=h_{b_k}\cdots h_{b_2}h_{b_1},
\hbox{ such that}\quad
t_{b_1}<t_{b_2}\cdots <t_{b_k},
\eqno(2.12)$$
and $\rho^J(d\om)$ is a probability measure, under which
$\om$ forms a Poisson process over $\B\times[0,\beta]$,
with the Poisson density $\prod_b J_b dt$.
Thus, $\om$ forms a random countable collection of
time-indexed bonds which occur independently in disjoint
regions of $\B\times[0,\beta]$.

The Poisson integral formula (2.11) offers a
non-commutative version of the familiar power series
expansion of the exponential function.

\noindent
{\bf 2.3 The Quasi-state decomposition}\nl

In a wide class of models there exists an orthonormal
basis $\{\ket{\alpha}\}$ of the Hilbert space of the
system such that
for all $\om$
$$
\bra{\alpha} \tprod_{(b,t)\in\om}h_{b}\ket{\alpha} \geq
0\quad .
\eqno(2.13) $$
For such models the Poisson integral formula (2.11)
provides a starting point for a quasi-state decomposition (Q-S decomposition)
 of the Gibbs state defined by
$$
\langle Q\rangle={\Tr e^{-\beta H} Q\over\Tr e^{-
\beta H}}\quad .
\eqno(2.14)$$
As we shall see, the condition (2.13) can be met, in suitable
bases, for both ferromagnetic and antiferromagnetic
models, and it does not  require the existence of a
basis in which all $h_b$ have only non-negative matrix
elements as was the case in various previous
approaches \tref{\CS,\Gin,\GM,\Ken,\KiTh,\Mata,\Matb,\Suto,\Tho,\Toth}
(for a treatment of a much wider class of interactions see \tref\AN ).

The Q-S decomposition resulting form  (2.13) takes the form:
$$
\langle \cdot\rangle=\int \mu(d\om) \langle
\cdot\rangle_\om
\eqno(2.15)$$
with
$$
\langle Q\rangle_\om ={\Tr K(\om) Q\over
\Tr K(\om)}
,\quad K(\om)=\tprod_{(b,t)\in\om}h_{b}\quad,
\eqno(2.16)$$
and
$$
\mu(\d\om)={\rho^J_{[0,\beta]}(\d\om)\Tr
K(\om)\over
\int \rho^J_{[0,\beta]}(\d\om)\Tr K(\om)}\quad .
\eqno(2.17)$$

Important features of the model are reflected in: i) the
structure of the quasi-states  $\langle \cdot\rangle_\om$,
and ii) the relative weights in the decomposition (2.15) of
different classes of $\om$.  To illustrate this, we now look
in detail at the spin-1/2 ferromagnetic and
antiferromagnetic models.

\noindent
{\bf 2.4 The spin-1/2 Heisenberg ferro- and antiferromagnet}\nl

The Heisenberg Hamiltonian is
$$\eqalignno{
H^{{AF\atop (F)}}&=  \textstyle{{+\atop (-)}} \tover12\sum_{<x,y>}
J_{x,y}\sig_x\cdot\sig_y  + \hbox{Const.}
&(2.18)\cr
&=-\sum_{<x,y>}J_{x,y}h^{{AF\atop (F)}}_{\{x,y\}}+ \hbox{Const.}&(2.19)\cr
}$$
where $\sig=(\sigma^1,\sigma^2,\sigma^3)$ are the usual Pauli
matrices
and the sum is over a set of pairs of sites. The signs are chosen such that we
can always assume that the coupling constants $J_{x,y}$ are non-negative.
The following choice of
$h_b$ permits us to cast both the ferromagnetic and the
antiferromagnetic Hamiltonians in the form of (2.10):
$$
h_b=\cases{T_{x,y}& ferromagnet\cr
2P^{(0)}_{x,y}& antiferromagnet\cr}
\eqno(2.20)$$
where $T_{x,y}$ and $P_{x,y}^{(0)}$ are the transposition and
singlet-projection operators acting in the Hilbert
space of the sites $x$ and $y$ (so, $T_{x,y}$ interchanges the states
at the sites $x$ and $y$). Use is made here of
the relations
$$
\sig_x\cdot\sig_y=2T_{x,y}-1=1-4 P^{(0)}_{x,y}
\eqno(2.21)
 $$

In discussing the matrix elements
$$
\langle \sigma\mid\tprod_{b\in\om} h_b\mid\sigma'\rangle
=
\sum_{\sigma_{t_1},\ldots,\sigma_{t_{k-1}}}
\bra {\sigma} h_{b_1}\ketbra{\sigma_{t_1}}{\sigma_{t_1}}
h_{b_2}\ket{\sigma_{t_2}} \cdots\bra {\sigma_{t_{k-1}}}
h_{b_k}\ket{\sigma'}
\eqno(2.22)$$
it is convenient to consider a space time picture in which
the RHS is viewed as a sum over paths in the spin configuration
space, with $\{\sigma^3_x\}$  defined at all times.  That
configuration is
piecewise constant, and the amplitude for the
process is determined by the matrix elements of the operators $\{h_b\}$.
Beyond this point the two cases need be discussed seperately.

\vfil\eject
\noindent
{\it 2.4.a: The ferromagnet ($H^{F}$).}\nl

In the ferromagnetic case, the $h_b$ are transpositions, which occur with
 amplitude 1.  Thus, the time-ordered product (2.12) consists of
a sequence of transpositions which result in a permutation which we denote
$\pi (\om)$. In particular, it is easily seen that
$$
\Tr K(\om) = 2^{l^F(\om)}
\eqno(2.23)$$
where $l^F(\om)$ is the number of cycles in the corresponding
permutation. (The factor 2 reflects the dimension of the
single-site Hilbert space.)   More generally,
quantities of the form $\Tr f(\{\sigma^3_x\})e^{-\beta H^F}$ are
described by the following construction - which is similar to but not quite
the same as the one presented in Section 2.1 for the Hamiltonian
considered there.

The paths $\sigma(t)$ which contribute are constrained, by the nature
of the trace, to have  $\sigma(t=0) = \sigma(t=\beta)$,
and thus are periodic in time.  The configuration $\om$ is
visualised, in a space-time graph, by a collection of ``horizontal'' bonds
connecting ``vertical'' lines indexed by the lattice sites. For each
$\om$,  the contributing spin configurations are obtained by
decomposing the graph into
a collection of loops. The ferromagnetic loops are obtained by replacing
the antiferromagnetic bonds in Figure 1 by ferromagnetic ones (see Figure 2).
The loop to which a point $(x,t)$
belongs is found by moving ``upward'' along the vertical line at that point
till a bond is reached. At a ferromagnetic bond $b=\{x,y\}$, the path crosses
from $x$ to $y$ and continues in the positive time direction.
This procedure is continued until the time
$t=\beta$ is reached at which point the path jumps to $t=0$ at the same site.
\ie $t=0$ and $t=\beta$ are identified.
The process is repeated until one comes back at the point $(x,t)$.

\vskip 5 truecm
\vfil

\noindent
{\bf Figure 2:} The graphical representation of the
ferro- and antiferromagnetic bonds used in the drawing of
the loop configuration $\om$.

\eject

The permutation $K(\om)$ and its cycle decomposition is easily
read from the above picture: $K(\om)$  takes $x$ into the site
where the loop drawn starting from $(x,0)$ returns, for the first time,
to $t=0$.

The discussion in Section 2.3 yields now the following proposition.

\iproclaim Proposition 2.2.
For the spin-$\tover12$ Heisenberg ferromagnet, the partition function
and the expectation values of observables generated by
$\{\sigma^3_x\}_{x\in\Gamma}$ are  given by:
$$
\Z^F_\beta=\Tr e^{\beta\sum_{<x,y>}J_{x,y}(T_{x,y}-1)}=
\int\rho^J_{[0,\beta]}(\d\om)\, 2^{\el^F(\om)}\quad,
\eqno(2.24)$$
and
$$
{\Tr f(\sigma^3)e^{-\beta H^F}\over
\Z^F_\beta}
=\int\mu_F(\d\om)\, E^F_\om(f)
\eqno(2.25)$$
where $\mu^F(\d\om)$ is the probability measure
$$
\mu^F(\d\om)=(\Z^F_\beta)^{-1}\rho^J_{[0,\beta]}(\d\om)2^{\el^F(\om)}
\eqno(2.26)$$
and the expectation functional $E^F_\om(f)$ is obtained
by averaging, with equal weights, over all the spins
configurations which take common values
($\eta_\gamma=\pm 1$) on the cycles $\{\gamma\}\in\om$,
corresponding to $K(\om)$:
$$
E^F_\om(f)=2^{-\el^F(\om)}\sum_{\eta_\gamma =\pm 1}f(\sigma(\eta))
\eqno(2.27)$$
\eproclaim

In fact, $E^F_\om(f)$ are quasi-states (adapted to
the algebra generated by $\{\sigma^3_x\}_{x\in\Gamma}$ according
to the definition in Appendix I).  In section 2.5,
below, we shall discuss the expectation values of other observables.
However, let us note here that two interesting choices for $f$ are:
$f_1(\sigma^3)=\sigma_x^3\sigma_y^3$ and
$f_2(\sigma^3)=\exp(\beta h \sum_{x\in\Gamma}\sigma_x^3)$.
Here, in contrast to our overall convention, $\sigma^3_x$ denotes a
Pauli matrix with eigenvalues $\pm 1$.  For
these functions
$E^F_\om(f)$ is given by
$$\eqalignno{
E^F_\om(f_1)&=\Ind{$(x,0)$ and $(y,0)$ are on the same
loop}&(2.28)\cr
E^F_\om(f_2)&=\prod_{\gamma\in\om}\cosh
(\vert\gamma\vert\beta h)&(2.29)\cr
}$$
where $\vert\gamma\vert$ denotes the number of times $\gamma$
intersects
the $t=0$ axis (so, $\vert\gamma\vert=$ the length of the cycle
$\pi_\gamma$).
In fact very similar expressions to
the ones above can be derived also for itinerant electron models.
The analogue of formula (2.29) for the Hubbard model was used by Aizenman
and Lieb in \tref\AL\ to derive a generalization of Nagaoka's Theorem to
finite temperatures.

\noindent
{\it  2.4.b: The antiferromagnet ($H^{AF}$).}\nl

A significant difference between the ferromagnetic interaction and the
antiferromagnetic one, which is seen already at the classical level,
is the possibility of ``frustration''.  Our analysis is restricted to the
frustration-free
case, which is characterized by the existence of a bipartite structure: the
lattice
$\Gamma$ decomposes into two sublattices, $\Gamma_A$ and $\Gamma_B$, with
the couplings between
two  sites restricted to be antiferromagnetic if the sites belong to distinct
sublattices, and, in more elaborate models in which both kinds of interactions
are present (see \tref\AN ), ferromagnetic within each sublattice.
For convenience we also define
$$
(-1)^{\vert x-y\vert}=\cases{+1&if $x$ and $y$ belong to the same sublattice\cr
-1&if $x$ and $y$ belong to distinct sublattices\cr}
\eqno(2.30)$$
As we shall see now, under the assumption of a bipartite structure,
the positivity condition (2.13) is satisfied even though not all the matrix
elements of the operators $K(\om)$ are positive.

In the computation of quantities of the form
$\Tr f(\{\sigma^3_x\})e^{-\beta H^{AF}}$
we need the
matrix elements of $h_b$ for the antiferromagnet:
$$
h_b=2P^{(0)}_b=\sum_{\alpha,\beta\in\{\tover12,-\tover12\}}
(-1)^{\alpha-\beta}\ketbra {\beta,-\beta}{\alpha,-\alpha}
\eqno(2.31)$$
Note the similarity between (2.31) and (1.2).

At this point we have a choice: the spin $1/2$
AF system can be discussed
in a form close  to that of the spin $1/2$ ferromagnet
or in a form which is suitable for the more general spin-S
models (with the Hamiltonian (2.1)).
In order to lay the grounds for a sequel to this paper,
where we encounter spin $1/2$
systems with mixed F and AF interactions, we shall present
the first option before treating the general spin case.

As in the ferromagnetic case, for a given $\om$
the allowed time dependent spin configurations $\{\sigma^3_x\}$
are piecewise constant (in time) and can change
only at pairs of sites where a bond occurs.
When a change occurs it is again a transposition.
However, there are the following differences from the ferromagnetic case:
\item{i)} there is a restriction that where a bond occurs
the two spins add up to zero (both before and after the event)
\item{ii)} at a given bond a transposition may or may not occur
\item{iii)} the amplitude for a given ``path'' is (-1) raised
to the number of
transpositions, \ie it equals the parity of the resulting permutation.

\iproclaim Lemma 2.3.
In a bipartite system, for each pair of configurations $\{\sigma, \sigma'\}$,
all the permutations which take $\sigma$ into $\sigma'$
and which can be written as products of transpositions exchanging
sites on different sublattices have a common parity,
denoted here $\sign(\sigma,\sigma')$, with
$$
\sign(\sigma,\sigma)=+1 .
\eqno(2.32)$$
\eproclaim
\proof:  Each transposition changes the number of positive spins on the
A-sublattice
by $\pm 1$.  Therefore, noting that the spin configurations take values
$\pm \tover12$, the parity of any admissible permutation is
$$
(-1)^{\sum_{x\in\Gamma_A}(\sigma_x-\sigma'_x)}  .
\eqno(2.33)$$
 \QED

We now derive the following formula for the matrix elements of the
operators $K(\om)$:
$$\eqalignno{
\langle \sigma\mid\tprod_{b\in\om} h_b\mid\sigma'\rangle
&=
\sum_{\sigma_{t_1},\ldots,\sigma_{t_{k-1}}}
\bra {\sigma} h_{b_1}\ketbra{\sigma_{t_1}}{\sigma_{t_1}}
h_{b_2}\ket{\sigma_{t_2}} \cdots\bra {\sigma_{t_{k-1}}}
h_{b_k}\ket{\sigma'}& \cr
&=
\sign(\sigma,\sigma')2^{l_0^{AF}(\om)}
{\rm\bf I}_{AF}[\sigma,\sigma'\mid\om]&(2.34) \cr
}$$
where $l_0^{AF}(\om)$ and ${\rm I}_{AF}[\sigma,\sigma';\om]$ are defined
graphically, in a
decomposition of the space-time graph into paths, which in many ways is similar
to the one seen in the ferromagnetic case.  The paths now come in two kinds:
open paths with end points of the form $(x,0)$ and $(x,\beta)$, and
closed paths which we call {\it internal loops\/} (see Figure 1).
The open paths are constructed by starting from any point $(x,t)$ with
$x\in\Gamma$
and $t=0$ or $\beta$, by moving in the vertical direction until a bond is met.
Upon the traversal of a bond, the orientation of the motion in time is
reversed.
The path stops upon reaching $t=0$ or $t=\beta$. The internal loops are
obtained in the same way, starting from any point
$(x,t)\in\Gamma\times [0,\beta]$ that does not already belong to an open path
and a loop is completed when the path comes back to its
starting point. $l_0^{AF}(\om)$ denotes the number of internal loops.
${\rm I}_{AF}[\sigma,\sigma';\om]$ is defined in terms of the open paths
which we interprete as imposing a pairing condition on the spin
configurations $\sigma=\sigma(0)$ and $\sigma'=\sigma(\beta)$.
The condition is that for any pair of points $\x=(x,t)$ and $\y=(y,t')$ with
$t$ and $t'$ either $0$ or $\beta$, that are the end points of a path in
$\om$, one has
$$
\sigma(t)_x\sigma(t')_y=(-1)^{\vert x-y\vert}=1-2\delta_{t,t'}
\eqno(2.35)$$
${\rm I}_{AF}[\sigma,\sigma'\mid\om]=1$ if (2.35) is satisfied for all open
paths in $\om$ and 0 otherwise.

Formula (2.34) is now a direct consequence of the graphical representation
of $\om$ and the definitions given above.

When computing $\Tr K(\om)$ one identifies $t=0$ and $t=\beta$. Then, all
paths are closed, \ie they are loops, and we write
$l^{AF}(\om)$ or $l_{\rm per}(\om)$ to denote the total number of loops in
$\om$.

The above discussion leads now directly to the representation given by
Proposition 2.1 for $S=1/2$.

\noindent
{\bf 2.5 Structure of the quasi-states}\nl

{}From the expressions in Proposition 2.1 and 2.2 it is obvious that the
quasi-states $\langle\cdot\rangle_\om$ depend only on the structural properties
of $\om$ revealed in the random loop picture of the configuration.
In fact, the only relevant property of $\om$ is how its random loops
link together sets of sites at $t=0$.
In both the ferro- and the antiferromagnet a quasi-state
$E_\om$ is uniquely determined by the permutation $\pi_\om$ of the sites
in $\Gamma$, which takes the site $x\in \Gamma$ to the site
$\pi(x)$ which is where the loop at $x$, starting off in the positive
time direction, intersects $t=0$ for the next time.
There is a one-to-one correspondence between the cycles in $\pi_\om$ and
the loops in $\om$ that intersect the $t=0$ hyperplane.
Let $\gamma_0$ denote such a generic cycle:
$$
\gamma_0=\pmatrix{x_1&x_2&\cdots&x_{r-1}&x_r\cr
x_2&x_3&\cdots&x_r&x_1\cr}
\eqno(2.36)$$
A first observation to make is then that in the functional
$E^\#_\om$, with $\#=F$ or $AF$, there are no correlations between the
spins on two sets of sites which support distinct cycles in the permutation:
$$
E^\#_\om(\prod_{\gamma_0} A_{\gamma_0})
=\prod_{\gamma_0}
E^\#_\om(A_{\gamma_0})
\eqno(2.37)$$
where $A_{\gamma_0}$ is an arbitrary operator acting on the sites
$\{x_1,\ldots,x_r\}$. Moreover, $E^\#_\om(A_{\gamma_0})$ depends on $\om$
only through $\gamma_0$ and we therefore might as well denote it by
$E^\#_{\gamma_0}(A_{\gamma_0})$.

As the $E^\#_\om$ are linear, they are
completely determined by their values on operators
of the form $A_{\gamma_0}=\sigma_{x_1}^{i_1}\cdots
\sigma_{x_r}^{i_r}$, where $i_j\in\{0,1,2,3\}$, and by $\sigma^0$ we
denote the identity operator. It is a straightforward excercise to
compute $E^\#_{\gamma_0}(\sigma_{x_1}^{i_1}\cdots
\sigma_{x_r}^{i_r})$ starting from the space-time picture:
we consider the spin to be a piecewise constant function along
$\gamma_0$ taking values $\sigma^3=\pm1$. Where there is a factor
$\sigma^3$ the expectation value picks up a factor
$\pm 1$ according to what the spin is at that point. $\sigma^1$
reverses the spin and $\sigma^2=i\sigma^1\sigma^3$. For the ferromagnet
the result is:
$$
E^F_{\gamma_0}(\sigma_{x_1}^{i_1}\cdots
\sigma_{x_r}^{i_r})=\tover12\{\bra{\tover12}\sigma^{i_1}\cdots\sigma^{i_r}
\ket{\tover12} + \bra{-\tover12}\sigma^{i_1}\cdots\sigma^{i_r}\ket{-\tover12}\}
\eqno(2.38)$$
For the antiferromagnet the only difference is that the spins now
have a staggered interpretation:
$$\sh^3_x=(-1)^{\vert x\vert}\sigma^3,\quad
\sh^2_x=\sigma^2,\quad
\sh^1_x=\sigma^1
\eqno(2.39)$$
This is equivalent to reversing the spin whenever the loop $\gamma_0$
changes its orientation with respect to the positive time axis, \ie whenever
it traverses an antiferromagnetic bond.
The relation (2.39) does of course not correspond to a unitary
transformation of the full algebra of observables of the system and is
possible only because we are not dealing
with true quantum states but only with quasi-states. A compact
expression for the antiferromagnetic functionals is:
$$
E^{AF}_{\gamma_0}(\sigma_{x_1}^{i_1}\cdots
\sigma_{x_r}^{i_r})=E^F_{\gamma_0}(\sh_{x_1}^{i_1}\cdots
\sh_{x_r}^{i_r})
=(-1)^{\#\{x_j\mid i_j=3\}
\cap \Gamma_A}
E^F_{\gamma_0}(\sigma_{x_1}^{i_1}\cdots
\sigma_{x_r}^{i_r})
\eqno(2.40)$$

We conclude this section by summarizing the results of Propositions
2.1 and 2.2 in the following way:
we found a representation of the ferro- and antiferromagnetixc
ground states as a convex combination of quasi-states which for the
ferromagnet are a partition of the lattice into subsets on which the spins
are locked together in a parallel state. For the antiferromagnet
the spins are rigidly correlated in a staggered manner. This picture
can be considered as a generalization of the VBS-states where
neigboring spins are paired into the singlet state (for a different
generalization see \tref\FNW ). In particular, depending on the
properties of $\mu(d\om)$, the states can also have long range order,
characterized by the fact that the clusters percolate throughout
the system.
Finally we want to remark that the quasi-states each have a much larger
symmetry group (which depends on $\om$) than the ground state itself.
Because there are no correlations between the sets of sites belonging to
different loops, the spins on the distinct cycles in $\pi_\om$ can
be rotated independently.

\noindent
{\bf 2.6  The SU($2S+1$)-invariant spin-s models with
interaction $-P^{(0)}$}\nl

We now turn to the Q-S decomposition of a generalization
of the spin-$\tover12$ Heisenberg ferro- and antiferromagnet to
arbitrary values of the spin. In particular this generalization includes
the one-dimensional antiferromagnets that are the main subject of this
paper.
Starting from the the Heisenberg Hamiltonians as they were written in
Section 2.4, we just replace $T_{x,y}$ and $2P^{(0)}_{x,y}$ by the
corresponding operators for a spin $S$ system: in the ``ferromagetic''
case the interaction $h_b^F=T_{x,y}$ interchanges
the states at the sites $x$ and $y$ and for the antiferromagnet $h_b^{AF}
=(2S+1)P^{(0)}_{x,y}$, where $P^{(0)}$ is given in (1.2).
It is obvious that both $h_b^F$ and $h_b^{AF}$ are
SU(2)-invariant. Due to the invariance of this interaction under parity
preserving relabelings of the $2S+1$ states, this SU(2)-symmetry is actually
embedded in a larger (for $S\geq 1$) SU($2S+1$)-symmetry.
The model with interaction $h_b^F$ is a degenerate
ferromagnet and its ground states are given by all permutation symmetric states
of the system. (The integrability of the one-dimensional models with
interactions $h_b^F$ and $-h_b^F$ was demonstrated in the well-known
work of Sutherland \tref\Sut ). The interaction
$h_b^{AF}$ is much more interesting and is the actual subject of the rest
of the paper. In the case of the one-dimensional lattice, we
recover the SU($2S+1$)-invariant model that was first studied by Affleck
\tref\Aff\ and also by Kl\"umper \tref\Klub\ and Batchelor and Barber
\tref\BBb .

As before, we can only deal with the non-frustrated case, and let $\Gamma_A$
and $\Gamma_B$ denote the two sublattices defining the bipartite structure
of the system.
The SU($2S+1$)-symmetry of the Hamiltonian with interaction (1.2) is
then represented by the fundamental representation on one sublattice and
the antifundamental representation on the other sublattice.

The matrix elements of $e^{-\beta H}$ and the partition function for this model
can be expressed using the same correspondence between configurations
$\om$ and sets of loops as for the spin-1/2 antiferromagnet. But now
each loop has to be decorated with a label $\alpha$ taking the $2S+1$ values
$-S,-S+1,\ldots,S-1,S$.
The spin configuration as a function of
$x\in \Gamma$ and $t\in [0,\beta]$, is now the following:
$$
S^3(x,t)=\cases{\phantom{-}\alpha \hbox{ if the label of the loop
is $\alpha$ and
$x\in\Gamma_A$}\cr
-\alpha \hbox{ if the label of the loop is $\alpha$ and
$x\in\Gamma_B$}\cr}
\eqno(2.41)$$

The proof of Proposition 2.1 now follows by the same arguments of
Section 2.4, taking into
account the correspondence between loop labelings and space-time spin
configurations given in (2.41).

In the one-dimensional case the loops can be interpreted as
the boundaries of the elements in a partition of the two-dimensional
space-time. As we will see
in the next section,
the weights with which these partitions occur are given
by the Gibbs weight for the associated configurations of
a Potts model at the selfdual point. This will enable us
to analyse the possible long-range order
in the ground state of these models.

\beginsection{3. Equivalence with the two-dimensional $(2S+1)^2$-state
 Potts models}

It will be natural to consider the spin chain with a priori different
coupling strengths for the  even and odd bonds. More generally, we are
concerned with the  Hamiltonian for a spin-$S$ chain given by:
$$
H^{AF}_{[L_-,L_+]}=-\sum_{x=L_-}^{L_+-1}
J_x h^{AF}_{\{x,x+1\}}
\eqno(3.1)$$
where $h^{AF}=(2S+1)P^{(0)}$ is defined in (1.2),
$J_x > 0$ and $L_-,L_+\in\Ir$, $L_-\leq 0<L_+$.

We shall now show that associated to the  geometric structure of Section 2 is
a Potts model, or rather a pair of dual Potts models
(the $A$- and the $B$-model).
In the translation invariant case ($J_x=J$ for all $x$) one arrives
at the Potts model at its self-dual point, where it is exactly solvable
\tref\Baxa . In this situation the equivalence was conjectured
by Affleck and established on the level of the
spectrum of the transfer matrices by Batchelor and Barber \tref\BBb\
and Kl\"umper \tref\Klu .

The Potts models are defined over a 1+1 dimensional lattice, $\Ir\times
\Rl$, in which one of the directions (corresponding to the ``time'' of
Section 2) is continuous.

To introduce the lattices on which the $A$- and $B$-Potts model variables
reside, we start from the space-time of the quantum spin chain embedded
in $\Rl^2$ and partition $\Rl^2$ into vertical strips of width $1$ which
we label alternatingly $A$ and $B$, with the strip  $0<x<1$ getting
the label $A$, as in Figure 1.
The Potts variables of the $A$-model reside on the vertical lines bisecting
the $A$-strips and the variables of the $B$-model are situtated on the
lines bisecting the $B$-strips.

For a volume $[0,T]\times [L_-,L_+]$, the Potts configurations of the $A$-model
are functions
$\xi(x,t)$, $x = \hbox{even} + \tover12 \in[L_-,L_+]$,  which are piecewise
constant in time and take values in $\{1,\ldots,q\}$.
In the time direction we always take the periodic boundary conditions.
 Other than that, there are two natural boundary conditions for a Potts model,
which are exchanged under the standard duality map: the {\it free} and the
{\it wired}, the latter corresponding to adding an extra strip to the left and
to the right of the volume
where the spins are required to assume a common value (we assume the convention
that the partition function includes also the sum over this common value).
As will emerge from our discussion, the relevant boundary conditions here
depend on the label of the strip along the boundary.  If it is $A$ then
the $A$-model gets the {\it free} b.c. and the $B$-model
the {\it wired} one, and otherwise it is the other way around.

The partition  function of the $A$-model (with the relevant boundary terms)
is given by
$$
\Z^{\rm Potts}_A=\int \rho_A^{J^V}(\d\om) \sum_\xi\phantom{}^\om
 \exp\{\sum_{{x=\hbox{\eightrm  even}+{1\over 2}\atop L_--1\leq x\leq L_+}}
\int_0^T \d t J_x^H (\delta_{\xi(x,t),\xi(x+2,t)} -1)\}
\eqno(3.2)$$
where $J^V=\{J^V_x\}$ and $\{J^H_x\}$ are sets of positive constants
(these are the ferromagnetic coupling constants in the vertical and horizontal
direction respectively and in which we have absorbed the inverse
temperature of the Potts model), $\rho_A^{J^V}$ is a product of independent
Poisson  point processes on the lines $\{x=2n+\tover12\}\times [0,T]$  with
intensity $J_x^V$,
and $\sum\phantom{}^\om$ denotes the sum over all configurations $\xi$
for which the discontinuities happen only at points $(x,t)$ in the
configuration
$\om$ of the Poisson process.
The sum in (3.2) should be interpreted as incorporating the
boundary condition convention explained in the
previous paragraph.

The Poisson measure  incorporates the interaction in the vertical
direction,
and it can be arrived at by way of a
continuum limit of ordinary discrete Potts models with vertical couplings
analogous to those seen in (3.2)
for the horizontal interaction.

For the
$B$-model we have an analogous expression with the condition
$\{x=$ even $+ \tover12\}$ replaced by $\{x=$ odd $+ \tover12\}$.

Potts models are conveniently studied via an embedding in the
random cluster model formulated by Fortuin and Kasteleyn \tref\FKa .
As we shall see, it is at that level
that the correspondence with the quantum spin chain is most explicit.
The theorem below and Theorem 7.2 present some key results which are derived
by that route.

\iproclaim Theorem 3.1.
Let the parameters of the finite quantum spin chain with
interaction $-P^{(0)}$ on the interval $[L_-,L_+]$,
and the ones of the $A$- and $B$-Potts models be related
as follows:
$$\eqalignno{
&J^H_x=(2S+1)J_{x+\tover12} ,\, J^V_x=(2S+1)^{-1}J_{x-\tover12} &\cr
&q = (2S+1)^2,\; \beta=T&(3.3)\cr
}$$
Then
\item{i)}
$$\eqalignno{
-{1\over\beta}\ln\Z^{(S)}&=-{1\over T}\ln\Z^{\rm Potts}_A
-{2S\over 2S+1}\!\sum_{{x={\rm  even}\atop L_-\leq x\leq L_+-1}}
J_x+2S\!\sum_{{x={\rm  odd}\atop
L_-\leq x\leq L_+-1}}J_x&\cr
&\qquad -{1\over T}\left(\tover12\{(-1)^{L_-}-(-1)^{L_+}\}-1
\right)\ln(2S+1)&(3.4)\cr
}$$
In particular the ground state energy per site
of the quantum chain equals the free energy per unit volume
of the associated Potts model up to a trivial constant.
\item{ii)} The Potts model has periodic boundary conditions in the
vertical direction and free boundary conditions in the horizontal direction.
\item{iii)} The distribution $\mu(\d \om)$ of the random loop
representation for the quantum chain as described in Section 2,
is identical to the distribution of the boundaries of the connected
clusters of the Potts model in the FK-representation (see below).
\eproclaim

It should be noted that  the coupling constants of both the $A$ and the
$B$ Potts model depend on all the coupling constants $\{J_x\}$ of
the quantum spin chain.  The relevant coupling
constants as defined in Theorem 3.1 are $J^V_x$ and $J^H_x$
with $x$=even $+\tover12$ for  $A$-model, and  $x$=odd
$+\tover12$  for the $B$-model.

In Section 7 we present some explicit relations between the
correlation functions of the quantum chain and  those of the Potts models,
which, in particular,  imply
 a non-perturbative result on the spectral gap in the quantum spin chains
with staggered couplings   (see Theorem 7.1).

We now turn to the proof of Theorem 3.1.

First we derive the  Fortuin-Kasteleyn
representation \tref{\For,\FKa} of the two-dimensional Potts models described
above,
by showing that the partition function and the probability measure
of the Potts model are equal to the ones of a random cluster model.
The random cluster model is
obtained by considering a configuration $\om=\om_A\times\om_B$
generated by the Poisson process $\rho^{J^V}_A(\d\om_A)
\rho_B^{J^H}(\d\om_B)$, as a partition of the union of the vertical
strips of type $A$ in the following way: a point $(x,t)\in\om_A$ is
considered as cutting the vertical strip with coordinate $x$ at the height
$t$ and a point $(x,t)\in\om_B$ connects the two pieces of $A$-strip
adjacent to the $B$-strip at coordinate $x$ at height $t$ (see Figure 1).
The partition function of the random cluster (RC) model is then given by:
$$
\Z^{\rm RC}=\int\rho^{J^V}_A(\d\om_A)\rho_B^{J^H}(\d\om_B)\; q^{C_A(\om_A
\times\om_B)}
\eqno(3.5)$$
where $C_A(\om_A\times\om_B)$ denotes the number of connected clusters
(consisting of pieces of strips of type $A$) in the configuration
$\om_A\times\om_B$. The equality of $\Z^{\rm Potts}$ and $\Z^{\rm RC}$ can be
derived straightforwardly by expanding the exponential in (3.2):
$$\eqalignno{
&\Z^{\rm Potts}_A=\int \rho_A^{J^V}(\d\om) \sum_\xi\phantom{}^{\om_A}
 \exp\{\sum_{x=2n+\tover12}
\int \d t J_x^H (\delta_{\xi(x,t),\xi(x+2,t)} -1)\}&\cr
&=\int \rho_A^{J^V}(\d\om) \sum_\xi\phantom{}^{\om_A}
\prod_x\sum_{n_x=0}^\infty
{e^{-\beta n_x}\over n!}\int\d t^{(x)}_1\cdots
\d t^{(x)}_{n_x} \delta_{\xi(x,t^{(x)}_1),\xi(x+2,t^{(x)}_1)}
\cdots  \delta_{\xi(x,t^{(x)}_{n_x}),\xi(x+2,t^{(x)}_{n_x})}&\cr
&=\int\rho^{J^V}_A(\d\om_A)\rho_B^{J^H}(\d\om_B)\;
\sum_\xi\phantom{}^{\om_A}\prod_{(y,t)\in\om_B}
\delta_{\xi(y,t),\xi(y+2,t)}&\cr
&=\int\rho^{J^V}_A(\d\om_A)\rho_B^{J^H}(\d\om_B)\;
 q^{C_A(\om_A\times\om_B)}=\Z^{\rm RC}&(3.6)\cr
}$$

In the sequel we shall write $\om$ for $\om_A\times\om_B$ and
$\rho^J(\d\om)$ or $\rho^{(J^V,J^H)}(\d\om)$ instead of
$\rho^{J^V}_A(\d\om_A)\rho_B^{J^H}(\d\om_B)$.

A by now standard and very convenient tool for the study of
the randon cluster model is provided by the FKG inequalities \tref\FKG .
Following is the FKG-structure which will be used to derive some of
the main results of this paper in Sections 4-7.

We define a partial order on the configurations $\om$ as follows:
$\om^\prime\preceq \om$ if the set of $A$-bonds in $\om$ {\it is contained in
\/}
the set of $A$-bonds in $\om^\prime$ and the set of $B$-bonds in $\om$
{\it contains\/}
the set of $B$-bonds in $\om^\prime$. (By $A$-bonds we mean bonds
occurring in an $A$-strip and analogously for $B$-bonds).
As before, let
$\el_{\rm per}(\om)$ denote the number of loops in $\om$, considered with
periodic
boundary conditions in time. We consider the loops as the boundaries of
a collection of connected subsets (connected clusters) of the plane.
Each such connected set consists of the vertical strips $n<x<n+1$
connected by horizontal bridges. All strips in a given cluster are either
of $A$ or of $B$ type. Let $C_A(\om)$ denote the number of connected
clusters of $A$ type and $C_B(\om)$ the number of clusters of $B$ type.
With these definitions one then has the
following obvious but crucial properties:
$N_A(\om)$ and $C_A(\om)$ are decreasing
functions of $\om$ and $N_B(\om)$ and $C_B(\om)$ are increasing, where
$N_C(\om)$ denotes the number of bonds in $\om$ which occur in the strips
of type $C$, $C=A,B$.

In the following proposition and throughout the rest of the paper,
$\rho^J(\d\om)$ denotes the Poisson measure on the configurations
$\om$ for the quantum spin chain on a finite interval, $[L_-,L_+]$,
containing $L_+-L_-+1$ sites, and at inverse temperature $\beta$ (see Section
2.1).
$J$ stands for the collection of coupling constants $\{J_x\}_{L_-\leq x L_+}$
which determine the intensities of the independent Poisson measures
for each bond. We will sometimes need to distinguish between the
coupling constants for even and odd $x$; we then use the notation:
$J=(\Je  ,\Jo  )$.

One then has the following relations:

\iproclaim Proposition 3.2.
With boundary conditions described above,
the following relations hold: for any $u>0$,
$$\eqalignno{
\rho^{(\Je  ,\Jo  )}(\d\om)\;
u^{\el_{\rm per}(\om)}
&=c_1 \rho^{(\Je  ,\Jo  )}(\d\om)\;
u^{C_A(\om)+C_B(\om)}&\cr
&=c_2 \rho^{(u^{-1}\Je  ,u\Jo  )}
(\d\om)\; u^{2C_A(\om)}&\cr
&=c_3 \rho^{(u\Je  ,u^{-1}\Jo  )}
(\d\om)\; u^{2C_B(\om)}&(3.7)\cr
}$$
where
$$\eqalignno{
c_1&=u^{-1}&\cr
c_2&= u^{\tover12 \{(-1)^{L_-}-(-1)^{L_+}\}-1}\exp\{\beta((1-u^{-1})
\sum_{x \hbox{\eightrm\ even}}J_x+(1-u)
\sum_{x \hbox{\eightrm\ odd}}J_x)\}&\cr
c_3&= u^{-\tover12 \{(-1)^{L_-}-(-1)^{L_+}\}-1}\exp\{\beta((1-u)
\sum_{x \hbox{\eightrm\ even}}J_x+(1-u^{-1})
\sum_{x \hbox{\eightrm\ odd}}J_x)\}&(3.8)\cr
}$$
\eproclaim
\proof:
\noindent
We start by reformulating the random loop picture as follows.
As before we consider the loops as being embedded in the plane
which we have divided into vertical strips labeled alternatingly $A$ and $B$
as in Figure 1. We can then associate with each configuration $\om$,
a set of $A$- and $B$-clusters as follows. When $\om$ contains an activated
bond in an  $A$-strip this bond is considered as forming a connection between
the
two $B$-strips to the immediate left
and right of it and as cutting through the $A$-strip to which it belongs,
and analogously for the bonds in a $B$-strip. Thus for each configuration
$\om$ we have obtained
a collection of clusters of
$A$ and $B$ type and such that two different clusters with a piece
of common border are of different type. $C_A(\om)$ and $C_B(\om)$
are the number of  $A$-clusters and $B$-clusters in $\om$, respectively.

We consider a finite interval of the chain of the form $[L_-,L_+]$, with
$L_-\leq 0 <L_+$. and such that the number of sites is even.

For concreteness, let us suppose first that the number of sites ($L_+-L_-+1$)
is even. Then, the boundary strips are either both of the $A$ or both of the
$B$ type; suppose they are of the $A$ type.
Recall that we then have free boundary conditions for the $A$-clusters
and wired boundary conditions for the $B$-clusters (wired independently
at the left and at the right boundary) in the space (horizontal) direction.
In the time (vertical) direction the boundary conditions are periodic.
Call everything connected to the leftmost $B$-cluster the ``outside''.
Then each loop in $\om$ unambiguously encloses
a domain and has the complement of that domain as its outside, \ie
every loop in $\om$ is the outer boundary
of exactly one domain and except for
the ``outside'' , each domain has a loop as its outer boundary. Therefore
the following relation holds
$$
\el_{\rm per}(\om)=C_A(\om)+C_B(\om)-1
\eqno(3.9)$$
Obviously we would have arrrived
at the same relation (3.7) if the boundary strips are of the $B$
type or if the number of sites is odd.
Denote by $N_A(\om)$ and $N_B(\om)$ the number of activated bonds that
occur in $\om$ in the $A$-strips and $B$-strips, repectively.
Then, with the boundary conditions described above,
the following Euler relation holds (see Lemma 3.3 below):
$$
C_B(\om)-C_A(\om)+N_A(\om)-N_B(\om)=\tover12 \{(-1)^{L_-}-(-1)^{L_+}\}
\eqno(3.10)$$
Using (3.9) and (3.10) we have:
$$\eqalignno{
\rho^J(\d\om)\; u^{\el_{\rm per}(\om)}
&=\rho^J(\d\om)\; u^{C_A(\om)+C_B(\om)-1}&\cr
&=u^{\tover12 \{(-1)^{L_-}-(-1)^{L_+}\}-1}\rho^J(\d\om)\;
u^{2C_A(\om)-N_A(\om)+N_B(\om)}&\cr
&=u^{-\tover12 \{(-1)^{L_-}-(-1)^{L_+}\}-1}\rho^J(\d\om)\;
u^{2C_B(\om)-N_B(\om)+N_A(\om)}&(3.11)\cr
}$$
The factors $u^{N_A(\om)}$ and $u^{N_B(\om)}$ can easily be absorbed in
the measure $\rho(\d\om)$ as a mere modification of the intensity and the
normalization of the Poisson process using the relation:
$$
\rho^J(\d\om)\,\lambda^{N(\om)}=
e^{\beta J(1-\lambda)}\rho^{\lambda J} (\d\om)
\eqno(3.12)$$
for any $\lambda>0$.  That completes the proof.
\QED

\noindent
{\bf Proof of Theorem 3.1:}\nl
In Proposition 2.1 we obtained the
following expression for the partition function of the \model :
$$
\Z^{(S)}=\int\rho^J(\d\om)\,
 (2S+1)^{\el_{\rm per}(\om)}
\eqno(3.13)$$
We apply the second equality of Proposition 3.2 with $u=2S+1$. Comparing
the result with (3.6) one then sees that up to a trivial constant
$\Z^{(S)}$ is equal to the partition function of the $A$-Potts model in
the Fortuin-Kasteleyn representation with $q=(2S+1)^2$. Taking the
logarithm one obtains i) of Theorem 3.1. In the same way one obtains
ii).
\QEDD

We still have to prove the particular form of the Euler relation that
was used in the proof of Proposition 3.2. We will use the following lemma
also to determine the self-dual point of the ``continuous time'' Potts
models.

\iproclaim  Lemma 3.3 (The Euler relation).
With the prescription of above one has:
$$
E(\om)\equiv C_B(\om)-C_A(\om)+N_A(\om)-N_B(\om)=\tover12
\{(-1)^{L_-}-(-1)^{L_+}\}
\eqno(3.14)$$
\eproclaim
\proof:
It is obvious that (3.14) holds for the configuration $\om$ that contains no
bonds at all. We now prove that (3.14) is valid in general by showing that
for any configuration $\om^\prime$ obtained from any other configuration $\om$
by
removing a bond, one has $E(\om^\prime)=E(\om)$.

Suppose for concreteness that the bond to be removed is in an $A$-strip
and call it $b$. We divide the neighbourhood of $b$ into four regions and
label them $A_1,A_2,B_1$ and $B_2$ as shown in
Figure 1. There are two possibilities:

\item{1)} in the cases where the domains $A_1$ and $A_2$ are connected with
the bond $b$ present, we must have that after the bond has been removed $B_1$
and
$B_2$ are no longer connected. So in this case removing $b$ leaves the number
of
$A$-clusters unchanged, but increases the number of $B$-clusters by
1. Hence indeed $E(\om^\prime)=E(\om)$.

\item{2)} if with $b$ present $A_1$ and $A_2$ are not connected, then there at
least one of these two $A$-clusters must be surrounded by a
$B$-cluster that contains both $B_1$ and $B_2$.  Therefore $B_1$
and $B_2$ will still be connected when $b$ is removed.  So, in this case
removing $b$ leaves the number of $B$-clusters unchanged and
decreases the number of $A$-clusters by 1. Again $E(\om^\prime)=E(\om)$.

\QED

The Potts models are exactly solvable at their self-dual point in the
thermodynamic limit.
For quantum spin chains this corresponds to the translation invariant case
where $J_x=J$ for all $x$, and taking the limit
$\beta\to\infty$. This relation follows from the next proposition.

\iproclaim  Proposition 3.4.
The relation
$$
{J^H\over J^V}=q
\eqno(3.15)$$
determines the self-dual point of the continuous time Potts model
with partition function (3.2).
\eproclaim
\proof:
The statement is a direct consequence of the Euler relation. In Figure 1 a
certain configuration $\om$ of cuts and bonds is depicted. The clusters
consist of the pieces of vertical strips connected by the horizontal
bonds in between. Recall that in the time direction the
boundary conditions are periodic. In the horizontal direction the clusters
at the boundary strips are subject to free boundary conditions.
We also added an extra strip without bonds to the left and to the
right of the interval. This is the situation where Lemma 3.3 holds and
the $A$ and the $B$ Potts model are then exactly each others dual.

Using Lemma 3.3 and (3.12), the expression (3.6) for the partition function of
the $A$ Potts model can then be transformed into the partition function
of the $B$ Potts model with new coupling constants:
$$\eqalignno{
\Z_A^{\rm Potts}&=\int\rho^{J^V}_A(\d\om_A)\rho_B^{J^H}(\d\om_B)\;
 q^{C_A(\om_A\times\om_B)}&\cr
&=q^{-\tover12\{(-1)^{L_-}-(-1)^{L_+}\}}\int\rho^{qJ^V}_A(\d\om_A)
\rho_B^{q^{-1}J^H}(\d\om_B)\; q^{C_B(\om_A\times\om_B)}&(3.16)\cr
}$$
As $\rho^{qJ^V}_A(\d\om_A)$ generates the horizonal and
$\rho_B^{q^{-1}J^H}(\d\om_B)$ the vertical bonds of the $B$-model,
the condition for self-duality is $J^H=qJ^V$.
\QED

Note that because of the relations between the parameters of the quantum
spin chain and the Potts models, self-duality of the Potts models
is equivalent to translation invariance of the spin chain.

We end this section by deriving an expression for the spin-spin correlation
function of the spin chain under consideration, in terms of the associated
random cluster model (3.6). This random
cluster model is a simultaneous realization of the $A$ Potts model
and its dual, the $B$ Potts model. As remarked before, the boundary conditions
are periodic in the vertical direction for both realizations,  but at the
left and right edges one has wired boundary conditions for one and free
boundary conditions for the other realization.

In terms of the random loop model the (imaginary time) spin-spin correlation
function is given by (see (2.9)):
$$
\langle S^3(x,t)S^3(y,s)\rangle
=(-1)^{x-y}C(S)\Prob{$(x,t)$ is connected by a loop to $(y,s)$}
\eqno(3.17)$$
In the random cluster model the loops are viewed as the boundaries
between two domains of opposite type. It is therefore clear that:
$$
\langle S^3(x,t)S^3(y,s)\rangle
=(-1)^{x-y}C(S)\mu(I_A(\x,\y)I_B(\x,\y))
\eqno(3.18)$$
where, for $C=A,B$, $I_C(\x,\y)$ is the indicator function of the event
that $\x_C$ and $\y_C$ belong to the same cluster.
Here $\x_C=(x_C,t)$, with $x_C$ defined by the requirements that
$\vert x-x_C\vert=\tover12$, and that $x_C$ belongs to  strip of type $C$.
Note that for any two space-time points
$\bx$ and $\by$, $I_A(\bx,\by)$ is an increasing and
$I_B(\bx,\by)$ a decreasing function.

We will often use the notation $\bx\sim\by$ to indicate the event
that $\bx$ and $\by$ are connected. If $\bx$ and $\by$ have integer
space coordinates (\ie they are on the  vertical lines) {\it connected} means
connected by a loop or a line in $\om$. If the space coordinates of
$\bx$ and $\by$ are not integer the notation means that they belong to
the same connected cluster. The event is empty
if the two points belong to strips of different type or if the space coordinate
of one of them is an integer and of the other one is not.

The individual probabilities $\Prob{$\x_A$ and $\y_A$ belong to the same
cluster}$, can be computed in the $q$-state Potts model (see the proof of
Proposition 3.2, where we applied the Euler relation). So,
$$
\langle S^3(\x)S^3(\y)\rangle(-1)^{x-y}
\leq C(S)\Probin{\hbox{\eightrm Potts}}{$\x_A$ and $\y_A$ belong to the same
cluster}
\eqno(3.19)$$
and if $\vert x-y\vert =1$ and $t=s$, we have
$$
\langle S^3(x)S^3(x+1)\rangle
=-C(S)\Probin{\hbox{\eightrm Potts}}{$x-\tover12$ and $x+\tover32$
belong to the same cluster}
\eqno(3.20)$$

\beginsection 4. Finite systems and the thermodynamic limit

As a first application of the equivalence with the Potts model obtained in the
previous section we now prove some preliminary properties of the \model .
Except when mentioned otherwise we only assume that the coupling constants
are strictly positive.

\iproclaim Theorem 4.1 (Finite systems).
For each finite interval $[L_-,L_+]$, the limit
$$
\langle\cdot\rangle_{[L_-,L_+]}=
\lim_{\beta\to\infty}\langle\cdot\rangle_{[L_-,L_+]}^\beta
\eqno(4.1)$$
 exists, both for the quantum spin
chain and for the associated random cluster model. Furthermore, if
the number of sites (\ie $L_+-L_-+1$) is even, the limiting quantum state
has $S_{\rm tot}=0$ and in the random cluster model there are no infinite
lines (almost surely).
\eproclaim

That the finite volume ground state of a finite chain of even length is unique
and is a spin singlet confirms the antiferromagnetic nature of the models.
The above theorem can be viewed as the analogue of the well known result of
Lieb and Mattis \tref\LM\ for the Heisenberg antiferromagnet for the class
of Hamiltonians under consideration here.

\iproclaim Theorem 4.2 (The thermodynamic limit).
\item{1)}  The finite volume ground states for the quantum spin chains and the
probability measures for the random cluster models converge to a well-defined
thermodynamic limit provided the parity of the boundary sites is
preserved; \ie for any local observable $Q$ the following limits exist:
$$\eqalignno{
\langle Q \rangle_{\rm even}&\equiv \lim_{L_-\to -\infty, L_- {\rm even}
\atop L_+\to +\infty, L_+ {\rm odd}} \lim_{\beta\to\infty}
\langle Q \rangle^{\beta}_{[L_-,L_+]}&(4.2)\cr
\langle Q \rangle_{\rm odd}&\equiv \lim_{L_-\to -\infty, L_- {\rm odd}
\atop L_+\to +\infty, L_+ {\rm even}} \lim_{\beta\to\infty}
\langle Q \rangle^{\beta}_{[L_-,L_+]}&(4.3)\cr
}$$
and with similar limits defining $\mu_{\rm even}(\d\om)$ and
$\mu_{\rm odd}(\d\om)$.
\item{2)}  The relation between the quantum states and the probability
measure of the random cluster model valid in finite volume (see Sections
2 and 3) persist in the infinite volume limit. In particular:
$$
\langle \es_x\cdot\es_y\rangle_{\rm even (odd)}=(-1)^{x-y}C(S)
\Probin{\mu_{\rm even (odd)}}{$(x,0)$ is connected by a loop to
$(y,0)$}
\eqno(4.4)$$
\item{3)}  For translation invariant couplings the states
$\langle \cdot\rangle_{\rm even}$ and $\langle \cdot\rangle_{\rm odd}$
 are translates of each
other, and each is invariant under translation of period two, as well as under
global spin rotations.
\eproclaim

\iproclaim Remark 4.3.
For translation invariant and staggered couplings the states
$\langle \cdot\rangle_{\rm even}$ and $\langle \cdot\rangle_{\rm odd}$
 are also ergodic
and weakly mixing \tref\Fur\ under even translations.
This fact follows from the clustering relation
and the clustering properties of the Potts model which are implied
by FKG arguments (see Section 7 (Theorem 7.2)).
\eproclaim

The rest of the section is devoted to the proofs of Theorems 4.1 and 4.2.

\noindent
{\bf Proof of Theorem 4.1}\nl
Since the arguments are fairly standard we shall be satisfied with an outline
of the proof.

The existence of the limiting quantum state for the finite
chains is trivial. For the random cluster model we can use the
one-dimensionality of the system and the Perron-Frobenius theorem.

If the number of sites in the spin chain ($L_+-L_-+1$) is even,
it follows that if there is one infinite line, then there must be at
least two infinite lines with opposite parity, \ie such that at any given time
the distance between the two lines is odd. In the random cluster picture
this means that there are two infinite clusters of the same color ($A$ or
$B$) which is impossible by the theorem of Burton and Keane \tref\BK .
Alternatively, one can show by a variational argument that maintaining
two infinite lines of opposite parity in a finite system costs a non-vanishing
amount of free energy per unit of imaginary time.

By construction the state of the spin chain is necessarily rotation
invariant and because there are no infinite lines one easily shows
that $\langle f(S^3_{\rm tot})\rangle= f(0)$ for any function $f$. Hence
the state must have $S_{\rm tot}=0$
\QEDD

\noindent
{\bf Proof of Theorem 4.2.}\nl
{}From the FKG structure defined in Section 3 it follows that the measures
$\mu_{[L_-,L_+]}(\d\om)$ form
a monotone sequence in FKG sense (see Appendix II) if $L_-$ and
$L_+$ are chosen such that the boundary strips are always of the
same type ($A$ or $B$). The standard argument then guarantees that the
thermodynamic limit converges (see \eg \tref\ACCN ).

That also for the quantum system the thermodynamic
limit converges and satisfies the relation (4.4) is not immediately evident
because local observables for the quantum chain are typically related
to probabilities of a non-monotone and non-local event, \eg
$$\eqalignno{
&\langle S^3_xS^3_y\rangle_{\rm even}&\cr
&=(-1)^{\vert x-y\vert}C(S)
\lim_{L_-\to -\infty, L_- {\rm even}
\atop L_+\to +\infty, L_+ {\rm odd}}
\Probin{\mu_{[L_-,L_+]}}{$(x,0)$ is connected by a loop to $(y,0)$}&(4.5)\cr
}$$
The event ``$(x,0)$ is connected by a loop to $(y,0)$'' is non-local and it is
non-monotone because
$$\eqalignno{
&\Probin{\hbox{\eightrm loops}}{$\x$ and $\y$ are connected by a loop}&\cr
&\qquad=\hbox{\rm Prob}_{\hbox{\eightrm random clusters}}\left(
{\hsize= 5.5cm\vbox to .5cm{\eightrm\noindent
$\x_A$ and $\y_A$ belong to the same cluster
and $\x_B$ and $\y_B$ belong to the same cluster}}\right)
&(4.6)\cr
}$$
which is not the probability of a monotone event since the connectivity
of the $A$ clusters is decreasing and the connectivity of the $B$ clusters
is increasing for the order structure at hand.
Therefore we need a separate argument to show that
$$\eqalignno{
&\lim_{L_-\to -\infty, L_- {\rm even}
\atop L_+\to +\infty, L_+ \rm odd}
\Probin{\mu_{[L_-,L_+]}}{$(x,0)$ is connected by a loop to $(y,0)$}&\cr
&\qquad\qquad\qquad =
\Probin{\mu_{\rm even}}{$(x,0)$ is connected by a loop to $(y,0)$}&(4.7)\cr
}$$
If (4.7) holds (and also its obvious extensions
to probabilities of more general connectivities of finite sets of sites),
the result follows by Proposition 2.1.

For convenience we introduce the following shorthand notation:
$$\eqalignno{
L&\equiv [L_-,L_+]&\cr
\Lambda_M&\equiv [-M,M]\times[-M,M]&\cr
L\to\infty&\equiv {L_-\to -\infty, L_- {\rm even} \atop L_+\to +\infty, L_+
{\rm odd}}&\cr
A_L&\equiv \Probin{\mu_{[L_-,L_+]}}{$x$ is connected by a loop to $y$}&\cr
A_{L,M}&\equiv \hbox{Prob}_{\mu_{[L_-,L_+]}}
(\hbox{\eightrm $x$ is connected to $y$
in the box $\Lambda_M$})&\cr
A_\infty&\equiv \Probin{\mu_{\rm even}}{$x$ is connected by a loop
to $y$}&(4.8)\cr
}$$
It is obvious that for each finite $L$
$$
A_L=\sup_M A_{L,M}
\eqno(4.9)$$
and also that
$$
0\leq A_L-A_{L,M}\leq
\hbox{\rm Prob}_{L}\left(
{\hsize= 5cm\vbox to .5cm{\eightrm\noindent
$x$ and $y$ are connected to $\Lambda_M^c$ but not
to each other by a loop inside $\Lambda_M$}}\right)
\eqno(4.10)$$
For $M$ fixed we take the $\limsup_{L\to\infty}$ of (4.10). For local
events the $\limsup$ is actually a convergent limit. Therefore:
$$\eqalignno{
0&\leq \limsup_{L\to\infty}A_L-
\Probin{\mu_{\rm even}}{$x$ and $y$ are connected inside $\Lambda_M$}&\cr
&\qquad \qquad \leq
\Probin{\mu_{\rm even}}{$x$ and $y$ are connected to $\Lambda_M^c$ but not
 inside $\Lambda_M$}&(4.11)\cr
}$$
The RHS is monotone decreasing in $M$. Taking the $\lim_{M\to\infty}$:
$$
0\leq \limsup_{L\to\infty}A_L-A_\infty\leq
\Probin{\mu_{\rm even}}{$x_A$ and $y_A$ belong to two distinct
$\infty$ clusters}
\eqno(4.12)$$
Since the RHS vanishes, the value of $\limsup_{L\to\infty}A_L$
is independent of the sequence of $L$'s and so the limit exists
and equals $A_\infty$.
\QEDD

\beginsection 5. Absence of N\'eel order

An infinite volume ground state of a quantum spin chain is said
to be N\'eel ordered if
$$
\liminf_r (-1)^r\langle \es_0\cdot\es_r\rangle >0
\eqno(5.1)$$
The aim of this section is to prove that in the \model\ N\'eel
order does not occur.
More specificly we prove the following theorem.

\iproclaim Theorem 5.1 (Absence of N\'eel order).
For translation invariant or staggered coupling constants the infinite volume
ground states  $\langle\cdot\rangle_{\rm even}$ and $\langle
\cdot\rangle_{\rm odd}$ satisfy:
$$
\lim_{r\to\infty}\langle \es_0\cdot\es_r\rangle_{\rm even
\atop (odd)}=0
\eqno(5.2)$$
\eproclaim

The absence of N\'eel order in the ground state of isotropic
antiferromagnetic quantum spin chains is believed to hold quite
generally, but a rigorous proof of this general fact is lacking.
More was done for the the
Heisenberg antiferromagnetic chains:  1) for $S=1/2$
the exact Bethe Ansatz solution shows no N\'eel order, and 2)
for general spin, an argument for the absence of N\'eel order
was given in \tref\PS\ on the basis of
a new correlation inequality (which is further discussed in \tref\Sha).

A key role for the argument is played by a result of
Gandolfi,  Kean, and Russo \tref\GKR , whose adaptation to our system takes
the following form.

\iproclaim Proposition 5.2.
For the wired state of the $A$ Potts model
$$\Prob{within\ a\ box\ of\ size\ $r$, $(0,0)$\ is\ surrounded\ by\ an\
$A$-connected\ path}\ \mathrel{\mathop{\kern0pt\longrightarrow}
\limits_{r\to \infty }}1
\eqno(5.3)$$
which holds regardless of $q$.
\eproclaim

It is easy to see that (5.3) is equivalent to the statement that there is
no $B$-percolation in the state obtained with the boundary conditions
which favor  $A$-connections.
The reason for that fact is that if
the $A$-clusters percolate then, then any finite region is encircled by
$A$-connected
closed paths which pervent $B$-percolation.
If there is no $A$-percolation in this $A$-preferred state then neither
do $B$-clusters percolate.
The proof of  \tref\GKR\ rests on  the planar
nature of the connectivity graph (\ie the nearest neighbor nature of the
interaction), its reflection symmetries,  and the FKG property.

\noindent
{\bf Proof of Theorem 5.1}

Theorem 4.2 permits us to express the expectation values of any local
observable
of the infinite  spin chain in terms of
the random  cluster measures
$\mu_{\rm even}(\d\om)$ and $\mu_{\rm odd}(\d\om)$.
The correspondence is identical to what  Proposition 2.1 provides
for finite systems.  In particular, by (2.9):
$$
(-1)^r\langle \es_0\cdot\es_r\rangle_{\rm odd} =
\Probin{\rm A-wired}{$(0,0)$ and $(r,0)$ are on the same loop}
\eqno(5.4)$$

As the loops are the boundaries of the connected clusters in the equivalent
random cluster model, when two sites $\bx$ and $\by$ are on the same
loop then both
$\bx_A\sim\by_A$ (in the $A$-sense) and $\bx_B\sim\by_B$
(in the $B$-sense).  Hence
the probability is bounded by
$$
\leq 1-\Prob{within\ a\ box\ of\ size\ $r$, $(0,0)$\ is\ surrounded\ by\ an\
$A$-connected\ path}\ \mathrel{\mathop{\kern0pt\longrightarrow}
\limits_{r\to \infty }}0
\eqno(5.5)$$
where the last step is by Proposition 5.2.
\QEDD

Let us remark
that assuming the validity of the exact results for the Potts
model, absence of N\'eel order is also implied by the bound
on the spin-spin correlation function
in terms of the truncated two-point function of the
Potts model, which is derived in Section 7.

\beginsection 6.  Dimerization versus Power Law Decay: a dichotomy

Despite the result of the last section, the models considered here may
exhibit symmetry breaking.  However, the symmetry is that of translation,
and the phenomenon is caused by dimerization of the spin chain.
The main result of this section is the following dichotomy: for the models
considered here, the ground state either dimerizes
(and exhibits spontaneous breaking of the translation invariance), or the
spin-spin correlations have a slow (non-exponential) decay, satisfying
$$
\sum_x \vert x\langle S^3_0 S^3_x\rangle\vert=+\infty   \quad.
\eqno(6.1)$$
Both possibilities occur (see the discussion at the end of Section 7).

A version of this dichotomy was first proved by Affleck and Lieb \tref\AffL ,
extending a result of Lieb, Schulz and Mattis
\tref\LSM.  They prove that the uniqueness of the ground state implies the
existence of low energy excitations (of the order of the inverse size of the
system).
The argument in \tref\AL\ applies only to half-integer SU(2)-spin
chains, and to SU($n$)-spin chains with a self-conjugate representation
acting at a site.   Thus, the domains of applicability of our analysis and
that of \tref\AffL\  have some partial overlap, but none includes the other.

We also show that, when translation
invariance is spontaneously broken, the two periodic states are distinguished
by
the nearest neighbour spin-spin correlation function.
In this sense (and more if one looks into the representation)
the states are dimerized.
The phenomenon can also be detected by the
long distance behavior of quantity:
$$
\O_{x,y}=e^{i{2\pi\over 2S+1}\sum_{u=x}^y S^3_u}
\eqno(6.2)$$
considered here only for   $x<y\in\Ir$ with $x-y$ odd.

Observables very similar to $\O_{x,y}$ have been used earlier
in studies of ground states of quantum spin systems
(see \tref{\AffL,\GA,\KT,\DNR}) and  in the computation of the
magnetization of the critical Potts model \tref\Baxb . In the second part
of this section we introduce the ``total spin on half of the infinite chain''
as
an operator in the GNS Hilbert space of the ground state. This quantity appears
to us as of more fundamental significance than the string observables
$\O_{x,y}$,
and the latter can be expressed in terms of it.  Moreover this new operator
serves as a dimerization order parameter which reveals more clearly the
detailed
nature of the dimer order.

\vfill\eject

\noindent
{\bf 6.1 The dichotomy}

Following is the main result of this section.

\iproclaim Theorem 6.1.
For the ground states of the translation invariant
\model ((3.1) with $J_x\equiv J$), the following dichotomy holds:
\item{--} either the translation symmetry is spontaneously broken in
the infinite volume ground states
\item{--} or the spin-spin correlation function decays slowly (non-exponential)
with
$$
\sum_x \vert x\langle S^3_0 S^3_x\rangle\vert=+\infty
\eqno(6.3)$$
In the first case, the symmetry breaking is manifested in the non-invariance
of the pair correlation:
$$
\langle \es_0\cdot\es_{1}\rangle_{\rm even}\neq
\langle \es_{1}\cdot\es_{2}\rangle_{\rm even}
=\langle \es_{0}\cdot\es_{1}\rangle_{\rm odd}
\eqno(6.4)$$
and also in the string order parameter:
$$
\eqalignno{
\lim_{N\to \infty}\langle \O_{0,2N-1}\rangle_{\rm even}
=\lim_{N\to \infty}\langle \O_{1,2N}\rangle_{\rm odd}&>0&(6.5)\cr
\lim_{N\to \infty}\langle \O_{0,2N-1}\rangle_{\rm odd}
=\lim_{N\to \infty}\langle \O_{1,2N}\rangle _{\rm even}&=0&(6.6)\cr
}$$
\eproclaim
\proof:
We again rely on  Theorem 4.2 to express the infinite volume state
in terms of
the random  cluster measures
$\mu_{\rm even}(\d\om)$ and $\mu_{\rm odd}(\d\om)$.
Since the {\it even} and the {\it odd} states are translates of
each other, it  suffices to consider
one of them, for concreteness say the odd state.

The physical origin of the dichotomy was outlined in the introduction.  The
random cluster representation permits us now to express these ideas
in a precise way.

First we note that  the result of \tref\GKR , on the absence of simultaneous
$A$ and $B$ percolation, implies that the configurations $\om$ contain no
infinite lines, \ie the lines (in space-time) along which the spins are
correlated
occur only in the form  of finite, non intersecting, closed loops.
The most important implication is that
 at any given time, the spins are locked into rigidly
correlated {\it even} clusters, with $\sum S^3=0$ within each cluster.

The alternative may now be posed as between the following two possibilities:
the number of loops surrounding each site  is either almost surely
finite, or it is almost surely infinite.  Equivalently, there either is
percolation
or no percolation (and if there is percolation then either $A$ or $B$
percolates, never  both, and hence the translation symmetry breaking).
The zero-one nature of these  probabilites
is due to the ergodicity of the measures under even translations (Remark 4.3).

In the absence of percolation, by the identity (2.9):
$$\eqalignno{
&\sum_{x\leq0,y\geq 1}\vert\langle S^3_x S^3_y\rangle\vert
=C(S) \sum_{x\leq0,y\geq 1}\Prob{$(x,0)\sim(y,0)$}&\cr
&\qquad=C(S) \ \E_{\rm odd}(\hbox{\#  connected pairs $\{(x,0),(y,0)\},
x\leq0, y\geq 1$})
= \infty&(6.7)\cr}
$$
where $\E_{\rm odd}$ is the expectation with respect to the probability
measure $\mu_{\rm odd}$.

That the alternative necessitates symmetry breaking can be seen in
different ways: i) via  the existence of either $A$ or $B$ percolation,  ii) by
the
distinction between the two sublattices in the values of
a topological index, or iii)  via the staggered values taken by the string
order
parameter.
More explicitly:

i)   If (6.1) fails then the origin is surrounded by a finite number of finite
loops.
The last of
those necessarily touches the infinite cluster. For the random cluster state
corresponding to the spin
 state $\langle\cdot\rangle_{\rm odd}$,
percolation is possible only for the $A$-cluster.  (The argument is presented
in
Section 5).  It then easily follows that
$$
\Probin{\rm odd}{ $x$ belongs to  an infinte cluster}
=\cases{m > 0&if $x$= even $+\tover12$\cr
 0&if $x$= odd $+\tover12$\cr}
\eqno(6.8)$$
A relevant question now is whether the
probabilities of these events, which are expressed purely in
terms of the configurations $\om$, can be expressed in terms of the
expectation values of some spin observables.
As we show below, the answer is -yes!

ii)  An alternative way to express the translation symmetry breaking,
though still at the level of the random cluster measure, is by considering
the topological index:
$$
w(\bx,\om)=(-1)^{\sum_{\gamma\in\om}{\bf I}(\gamma {\rm\ encloses\ } \bx)}
\eqno(6.9)$$
The index is defined only if the number of loops is finite, and it is easy to
see
 that, when defined, its values alternate (with the overall phase dependent
on $\om$).

iii)  We now show that the string order parameter (6.2), which
 is an observable
of the quantum spin chain, directly detects cluster
connectivity (as opposed to loop connectivity which determines
the spin-spin correlation).   By an application of (2.6-7):
$$
\langle \O_{x,y}\rangle_{\rm even(odd)}=\int\mu_{\rm even(odd)} E_\om(\O_{x,y})
\eqno(6.10)$$
with
$$
E_\om(\O_{x,y})=\Ind{each  loop of $\om$ intersects $[x,y]$ an
even number of times}   \quad .
\eqno(6.11)$$
We use here the observation that the net flux of any simple loop
through an interval is either zero or $\pm 1$.  If $\om$ has
a loop with a non zero flux through the interval $[x,y]$ then the
conditional average of $\O_{x,y}$, over the consistent spin configurations,
vanishes.  In the other case,   $\sum_{u=x}^y S^3_u\equiv 0$.

Hence, the string order
parameter can be given a neat geometrical
meaning:
$$\eqalignno{
\langle\O_{x,y}\rangle&=\Prob{all loops intersect $[x,y]$ an
even number of times}\cr
&=\Prob{any loop that encloses $(x-\tover12,0)$ also
encloses $(y+\tover12,0)$ and vice versa}\cr
&=\Prob{$(x-\tover12,0)$ and $(y+\tover12,0)$
belong to
the same cluster}&(6.12)\cr
}$$
It immediately follows that $\langle\O_{x,y}\rangle=0$ if $y-x-1$ is odd.

For the {\it odd} boundary conditions the percolating cluster can only be
of the $A$-type and it is unique.  It follows that
$$\eqalignno{
\lim_{N\to\infty}\langle\O_{x,x+2N+1}\rangle_{\rm odd}
&=\lim_{N\to\infty}\Probin{\mu_{\rm odd}}{$(x-\tover12,0)$
and $(x+2N+\tover32,0)$ belong to the $\infty$ cluster}\cr
&=\cases{m^2& for $x$  odd\cr
                 0  &  for $x$  even\cr}&(6.13)\cr
}$$
Equation (6.13) is an explicit proof that the alternative to (6.7)
is symmetry breaking.

Going beyond the last statement, we can also show that the
translation symmetry breaking is  necessarily manifested in
the nearest neighbour correlations.   By (4.4):
$$\eqalignno{
\langle S_x^3 S^3_{x+1}\rangle_{\rm odd}&=
-C(S)\Probin{\mu_{\rm odd}}{$x-\tover12\sim x+\tover32$}&\cr
&=\cases{\int\mu_{\rm odd}(\d\om)I_A(x,x+1)& if $x$ is odd\cr
         \int\mu_{\rm odd}(\d\om)I_B(x,x+1)& if $x$ is even\cr}&\cr
&=\cases{\int\mu_{\rm odd}(\d\om)I_A(x,x+1)& if $x$ is odd\cr
         \int\mu_{\rm even}(\d\om)I_A(x,x+1)& if $x$ is even\cr}&(6.14)\cr
}$$
where we also used the duality relation.

When there is percolation, $A$-clusters percolate
 in the state $\mu_{\rm odd}$ but not in the state $\mu_{\rm even}$.
The fact that this difference is then  detected also at the
level of the nearest - neighbor connections is
an implication of
the general criterion of the  Rising Tide Lemma, which is
derived here  in Appendix II. (In the terminology explained there,
$I_C(x,y)$ (with $C=A,B$) are {\it strictly\/} monotone functions.)
\QED

\noindent\nl
{\bf 6.2  The dimerization order parameter }\nl

More can be said about the dimerized state in algebraic terms,
by refering to the Hilbert space associated with the ground states
\even\ and \odd\ via the GNS construction.

In physical terms, when it is correct to view the spins as organized
into neutral clusters it is natural to talk about the total excess
spin to the right of $x$, i.e., the total spin in half
of the chain.  This observation explains the following claim,
which can be derived  within our representation of the states
\even\ and \odd .

\iproclaim Claim 6.2.
Under the condition:
$$
\sum_x \vert x\langle S^3_0 S^3_x\rangle\vert<\infty
\eqno(6.15)$$
(i.e., the opposite of (6.1)) the folowing limits exist
$$
\hat S_x^3=\mathop {\lim}\limits_{\varepsilon \downarrow 0}\
\sum\limits_{y>x} {e^{-\varepsilon |y-x|}\  S_y^3}=
\mathop {\lim}\limits_{\varepsilon \to 0}\ \hat S_x^3\;(\varepsilon \;) \ \ ,
\eqno(6.16)$$
 in the sense of
strong resolvent convergence of operators in the GNS Hilbert space associated
with either \even or \odd.
\eproclaim

In terms of the random cluster representation:

$$
\hat S_x^3=\  \sum\limits_{y>x} {\  S_y^3}\ \Ind{the loop of $y$
intersects $(-\infty ,x]$\ }
\eqno(6.17)$$

We omit here the proof, except for the comment that  what is proven
explicitely is  the convergence of the quantities:
$$
\langle A\  e^{it\hat S_x^3\;(\varepsilon \;)}\  B\rangle _{\rm even(odd)}
\eqno(6.18)$$
for all strictly local spin observables.  Equation  (6.14)
is a natural condition for both the existence
of the limiting operator $\hat S_x^3$ and for the proof of the convergence.
The first  statement is the simpler task, e.g., it is easy to see that:
$$\langle \  | {\hat S_x^3} |^2\  \rangle _{\rm even(odd)}\le \
 \sum\limits_{y>0} {\  |  y\langle S_0^3S_y^3\rangle |}
\eqno(6.19)$$

The operator $\hat S_x^3$ is the third component of a vector (under spin
rotations),
with $\hat S_x^1$ and $\hat S_x^2$ defined analogously.  Using the
the invariance of the states:
$$
\langle \  e^{{{2\pi i} \over {2S+1}}\hat S_x^3}\  \rangle _{\rm even(odd)}\
=\  \langle \  P_{|\hat S_x^{}|^2=0}\  \rangle _{\rm even(odd)}
\eqno(6.20)$$
where $ P_{|\hat S_x^{}|^2=0}$ is the orthogonal projection onto the subspace
with $|\hat S_x^{}|^2=0$.

The operators thus  constructed permit to express the string quantity as
follows:
$$
\O_{x,y}= e^{{2\pi i\over 2S+1}(\hat S_{x-1}^3-\hat S_y^3)}\ \ \hbox{ for }
y>x\in\Ir
\eqno(6.21)$$
Under the condition (6.15), the formula (6.17) for  $\hat S_x^3$, implies
the clustering behavior:
$$
\langle\O_{x,y} \rangle_{\rm even(odd)} - \langle P_{|\hat
S_{x-1}^{}|^2=0}\rangle
\langle P_{|\hat S_{y}^{}|^2=0}\rangle \to 0
\eqno(6.22)$$
for $|x-y|\to\infty$.
Hence, by comparison with (6.13) or directly from (6.17),
$$
\langle \  P_{|\hat S_0^{}|^2=0}\  \rangle _{\rm odd}=\  m\  \  ,\qquad
  \langle \  P_{|\hat S_1^{}|^2=0}\  \rangle _{\rm odd}=\  0
\eqno(6.23)$$

Thus, the rotation of all the spins to the right of an odd site,
by an angle $2k\pi /(2S+1),\, k=1,\ldots,2S$, produces a state orthogonal to
the ground state \odd .  The orthogonality expresses the fact  that such a
rotation
will necessarily break an existing bond.  That is not the case for rotations
of the spins to the right of an even site since, within the odd state, there
is a positive probability that none of the clusters within which the spins are
correlated are broken by this division.

\iproclaim Remark 6.3.
In the above discussion we used percolation ideas to relate the
failure of (6.1) to
the positivity of $m$.  For half integer spins,
 that can be replaced by an alternative argument based on the operators
 $\hat S_x^3$, which
shows how remarkable is the fact of their existence.
For half integer spins
$$
E_w(\  P_{|\hat S_x^{}|\in \Ir}\  )\  =\  \Ind{ $w(x,\omega )=+1$ }\  =\  0,1
\eqno(6.24)$$
where the important observation is  the 0-1 property.
Therefore, due to the ergodicity of the state  $\mu_{\rm odd}(\d\om)$
under the even translations, for each $x$:
$$
\langle \  P_{|\hat S_x^{}|\in Z}\  \rangle _{\rm odd}=\  0,\  1\
\hbox{depending on the parity of } x
\eqno(6.25)$$
Since,
$$
\hat S_x^3 = \hat S_{x+1}^3 + S_{x+1}^3 \ \  ,
\eqno(6.26)$$
the projections obey
$$
P_{|\hat S_x^{}|\in \Ir} = 1\  -P_{|\hat S_{x+1}^{}|\in \Ir}
\eqno(6.27)$$
That directly implies the lack of translation invariance, in the following
explicit form:
$$\eqalignno{
\langle e^{2\pi i \hat S_x^3}\rangle_{\rm odd}
&=\langle P_{\vert\hat S_x\vert \in\Ir}\rangle_{\rm odd}
-\langle P_{\vert\hat S_x\vert \in\Ir+\tover12}\rangle_{\rm odd}&\cr
&=(-1)^{x}&(6.28)\cr
}$$
where the overall phase was determined by parity considerations.
\eproclaim

This line of reasoning
 is reminiscent of the structural proof, by  Aizenman and Martin
 \tref\AM,  of symmetry breaking
in one dimensional Coulomb systems.
It may be noted that a string quantity related to the exponent seen in
(6.28) was used in the argument of \tref\AffL, which was also
restricted to half integer spins.

The above argument
expresses a different mechanism for dimerization
 than the one used in the proof of
Theorem 6.1.  The restriction to half-integer spin is compensated by the fact
that
(6.28) can be extended to a different class of (frustration free)
Hamiltonians for which the percolation picture is not valid, where
the result does hinge on the parity of $2S$.

\beginsection 7. Decay of correlations in the \model

Some elementary properties of the infinite volume limit of the states
$\langle\cdot\rangle_{\rm even}$ and $\langle\cdot\rangle_{\rm odd}$
were discussed in Theorem 4.2.
We now present sufficient conditions for the exponential decay of correlations
in
these states.  The main result is:

\iproclaim Theorem 7.1.
If either the A or the B Potts model associated with
the ground state of the Hamiltonian
$$
H=-\sum_x J_x P^{(0)}_{x,x+1}  , \qquad J_x >0 \hbox{ for all } x,
\eqno(7.1)$$
has an exponentially decaying truncated pair correlation
function (of the Potts variables),
then the limiting states $\langle\cdot\rangle_{\rm even}$
and $\langle\cdot\rangle_{\rm odd}$
exhibit exponential decay of correlations of local observables,
and a non-vanishing spectral gap above the ground state energy.
\eproclaim

For a more explicit statement we introduce the following terminology.
Let $Q$ be a local observable of the form
$$Q=e^{-tH}\,\prod\limits_{i=1}^k {S_{x_i}^{\alpha _i}}\;e^{tH}
\eqno(7.2)$$
(any local observable can be written as a finite sum of such products).
Then $\supp Q$ (the support of $Q$) is the set of space-time points
$\{(x_i,t)\}$. We also define $\overline{\supp}Q$ to be the minimal
interval (consisting of points at time $t$ with integer space
coordinate) containing
$\supp Q$.

The truncated correlation of two local
observables $Q$ and $Q^{'}$, is defined as $\langle Q\,;Q^{'}\rangle=
\langle QQ^{'}\rangle-
\langle Q\rangle\langle Q^{'}\rangle$. For the $A$ Potts model the
 truncated
two-point function, in a state which is symmetric under permutations
of the $q$ values of the spin, has the random cluster interpretation
$$\eqalignno{
\tau_A(\bx,\by)&=\langle \xi_{\bx_A}\xi_{\by_A}\rangle
-\langle \xi_{\bx_A}\rangle\langle\xi_{\by_A}\rangle&\cr
&={c(q)\over q-1}(q\langle \delta_{\xi_{\bx_A},\xi_{\by_A}}\rangle
-1)&\cr
&=c(q)\Prob{$\bx_A$ and $\by_A$ belong to the same connected cluster}&(7.3)\cr
}$$
where
$$
c(q)={1\over 12}(q^2-1)
\eqno(7.4)$$
with a similar relation holding for the $B$ Potts model.

\iproclaim Theorem 7.2.
Let $\langle\cdot\rangle$ denote the expectation in the ground state of
a finite chain containing an even number of sites, or in one of the limiting
states $\langle\cdot\rangle_{\rm even}$ or $\langle\cdot\rangle_{\rm odd}$ of
Theorem 4.2.
Then, for any pair of local observables of the quantum system
of the form (7.2),
$$
\vert\langle Q\,;Q^{\prime}\rangle\vert\leq
C_Q C_{Q^{\prime}} \sum_{\by\in\overline\supp Q\atop
\bz\in\overline\supp Q^\prime}\min_{C=A,B}\tau_C(\by,\bz)
\eqno(7.5)$$
where $C_Q$ and $C_{Q^{\prime}}$ are invariant under space-time
translations of the observables.
The relations between the
coupling constants of the quantum spin chain and the Potts models are
given in Theorem 3.1.
\eproclaim

The minimum over the two Potts models in (7.5) is important since
it expected that in any situation one of
the correlation functions vanishes, as $(\bx-\by)\to\infty$.
If that is indeed the case, then
the implication is that the limiting states
$\langle\cdot\rangle_{\rm even}$ and $\langle\cdot\rangle_{\rm odd}$ are always
clustering and hence pure phases. In the case of translation invariant or
staggered
couplings we have the following remark.

\iproclaim Remark 7.3.
The following inequality is obvious:
$$
\tau_C(\bx,\by)\leq c(q)\hbox{\rm Prob}\left({\hsize= 5.7cm\vbox to
.5cm{\eightrm
\noindent
the connected cluster of $\bx_C$ reaches beyond
a box of size $2\Vert\bx_C-\by_C\Vert$ centered at $\bx_C$}}\right)
\eqno(7.6)$$
Therefore, whenever one can show that there is no simultaneous percolation
of the $A$- and $B$-clusters, one  also obtains
$$
\lim_{\Vert\bx-\by\Vert\to\infty}\min_{C=A,B}\tau_C(\bx,\by)=0
\eqno(7.7)$$
In the case of translation invariant or staggered coupling constants the
argument of \tref\GKR\ applies,  but we expect that simultaneous $A$ and
$B$ percolation is absent under much weaker conditions.
\eproclaim

The bound (7.5), and some of the arguments used in its derivation are
reminiscent of the estimate derived in \tref\AKN\ for the quantum Ising
model in transverse field.  However, the case considered here is
less direct. From (3.18) it is clear that the spin-spin correlation
function of the quantum chain is not equal to the truncated two-point
function of the associated Potts model. Nevertheless the decay rates of the
two are equal (Theorem 7.6).

At the end of this section we discuss some implications for
models with alternating coupling constants.

There are three steps to the proof of Theorem 7.2:

1)  When the truncated correlations of quantum spins are transcribed
in terms of the Potts model, we obtain two distinct contributions.
The first is easily bounded in terms of the random cluster model's
connectivity function, and the second is a truncated correlation
function of suitable observables of the random cluster model.

2) Using a general domination principle for FKG measures,
the latter correlation function is bound in terms of
four point functions of the form
$$
\langle I_{C_1}(\bx,\by) ; I_{C_2}(\bu,\bv)\rangle
\eqno(7.8)$$
where $C_i$ are either A or B, and $I_C(\bx,\by)$
are indicator functions for the events that
$\bx_C$ and $\by_C$ are in the same connected $C$-cluster
(for the definition of $\bx_C$ see at the end of Section 3).

3)  It is shown that for any combination of $C_1$ and $C_2$
the truncated correlation (7.8) is bounded in terms
of  the connectivity function of the $A$-model.  By symmetry, the
same is true for the $B$-model (in any situation, only one of these
bounds will have a non-trivial consequence).  The argument is based
on a combination of the  FKG inequality and the Markov property
of the random cluster measure.

The argument relies on two auxilliary results. The first one,
used in step 2,
is the following domination principle for correlation
functions with respect to an FKG measure.

\iproclaim Lemma 7.4  (\tref\New ).
For a pair of functions $f$ and $g$, let $F$ and $G$
be two monotone functions with which  $f+F$ and $g+G$ are
increasing and  $f-F$ and $g-G$ are decreasing.
Then the truncated correlations with respect to
any measure with the FKG property satisfy:
$$
\vert\langle f\,; g\rangle\vert\leq \langle F\,; G\rangle
\eqno(7.9)$$
\eproclaim

The proof is elementary: it consists of two linear combinations
of the four correlation inequalities resulting from the monotonicity
of $f\pm F$ and $g\pm G$.

The next Lemma is needed for step 3. It provides an upper bound
for the truncated correlation of two monotone increasing functions
of the random cluster model.

\iproclaim Lemma 7.5.
Let $I_1$ and $I_2$ be two monotone increasing events for the random
cluster model, which are
determined by the $A$-connected clusters of two non-random sets
$D_1$ and $D_2$. Then:
$$
0\le \left\langle I_1\,;I_2\right\rangle \le
\Prob{$D_1$ and $D_2$ are $A$-connected}
\eqno(7.10)$$
\eproclaim
\proof:
The key observation is the following inequality which results
from the combination of the Markov structure and the FKG property
of the random cluster measure:
$$
\E[I_2\mid I[D_1\not\sim D_2] I_1]\leq\langle I_2\rangle
\eqno(7.11)$$
The reason for this inequality is that in the complement of
$C_A[D_1]$, the $A$-connected cluster of $D_1$, the system can
be considered as having free boundary conditions on the
boundary. These boundary conditions mask
the positive event $I_1$ which is determined within $C_A[D_1]$.

The above reasoning was first employed in
\tref\CNPR\ for the Ising model and similar arguments have
later been used for a variety of other applications.

Using (7.11) and the
trivial bound $\vert I_1\vert\leq 1$, one then obtains:
$$\eqalignno{
{\langle I_1I_2\rangle\over \langle I_2\rangle}=\E[I_1\mid I_2]
&=\E[I_1 \mid I[D_1\sim D_2]I_2]\E[I[D_1\sim D_2]\mid I_2]&\cr
&\quad +\E[I_1 \mid I[D_1\not\sim D_2]I_2]\E[I[D_1\not\sim D_2]\mid I_2]&\cr
&\leq \E[I[D_1\sim D_2]\mid I_2] +\langle I_1\rangle&(7.12)\cr
}$$
Thus,
$$
\langle I_1\rangle\langle I_2\rangle\leq
{\langle I_1I_2\rangle}\leq\langle I_1\rangle\langle I_2\rangle
+\E[I[D_1\not\sim D_2]]
\eqno(7.13)$$
where the first inequality is just by FKG.  That
implies the bound stated in the Lemma
for the truncated correlation $\langle I_1 ;I_2\rangle$.
\QED

\noindent
{\bf Proof of Theorem 7.2:}\nl
For convenience, we first carry the analysis for observables which are
products, of the form (7.2), of only $S^1$ and $S^3$ variables at
non-coincidental sites, and break the proof into the steps described
above.

1)  In terms of the random cluster representation of the spin chains:
$$
\langle Q\,;Q^{\prime}\rangle =
\int \mu(\d\om) E_\om(QQ^{\prime})-\left(\int \mu(\d\om) E_\om(Q)\right)
\left(\int \mu(\d\om^{\prime}) E_{\om^{\prime}}(Q^{\prime})\right)  \qquad.
\eqno(7.14)$$
Therefore
$$\eqalignno{
\vert\langle Q\,;Q^{\prime}\rangle\vert
&\leq \left\vert\int \mu(\d\om) E_\om(Q Q^{\prime})
-\int \mu(\d\om) E_\om(Q)E_\om(Q^{\prime})\right\vert&(7.15)\cr
&\quad +\left\vert\int \mu(\d\om) E_\om(Q)E_\om(Q^{\prime})
-\left(\int \mu(\d\om) E_\om(Q)\right)\left(\int \mu(\d\om^{\prime})
E_{\om^{\prime}}
(Q^{\prime})
\right)\right\vert&\cr
}$$
The first term on the right side can be interpreted as the average
over $\omega$ of the
truncated correlation {\it within} the quasi states $E_\om$.
The second term is a
truncated correlation function of the random cluster model.

To estimate the first term we note  the factorization property (see (2.37))
$$\eqalignno{
E_\om(QQ^{\prime})=E_\om(Q)E_\om(Q^{\prime})&\quad\hbox{ if $\om$
does not contain a loop }&(7.16)\cr
&\quad\hbox{ connecting $\supp Q$ and $\supp Q^{\prime}$}&\cr
}$$
Using also the bound $\vert E_\om(Q)\vert\leq S^{\vert\supp Q\vert}$,
one  has
$$\eqalignno{
&\left\vert\int \mu(\d\om) E_\om(Q Q^{\prime})
-\int \mu(\d\om) E_\om(Q)E_\om(Q^{\prime})\right\vert\quad&\cr
&\qquad\leq
S^{\vert\supp Q\vert +\vert\supp Q^{\prime}\vert}
\sum_{\bx\in\supp Q\atop \by\in \supp Q^{\prime}}
\Prob{$\bx$ and $\by$ are connected by a loop}&(7.17)\cr
}$$
When $\bx$ and $\by$ are connected by a loop then both
$\bx_A\sim\by_A$ and $\bx_B\sim\by_B$,
 where $\bx_C\sim\by_C$ is our notation for the event that
$\bx_C$ and $\by_C$ belong to the same $C$ cluster.
Thus we  can continue the bound as:
$$\eqalignno{
&\leq S^{\vert\supp Q\vert +\vert\supp Q^{\prime}\vert}
\sum_{\bx\in\supp Q\atop \by\in
\supp Q^{\prime}}\min_{C=A,B}\Prob{$\bx_C\sim\by_C$}&\cr
&\leq
c(q)^{-1}\quad S^{\vert\supp Q\vert +\vert\supp Q^{\prime}\vert}
 \sum_{\by\in\supp Q\atop
\bz\in\supp Q^\prime}\min_{C=A,B}\tau_C(\by,\bz)&(7.18)\cr
}$$

2)  In order to estimate the second term in the RHS of (7.15)
we invoke Lemma 7.4.
For the functions
$$
f(\omega )=E_\omega (Q),\ \ \ \ \ g(\omega )=E_\omega (Q') \ \ ,
\eqno(7.19)$$
we take
$$
F(\om)= 2S^{\vert\supp Q\vert} \left(\sum_{\bx,\by\in\supp Q}
I_A(\bx,\by)-I_B(\bx,\by)\right)
\eqno(7.20)$$
and
$$
G(\om)= 2S^{\vert\supp Q^\prime\vert} \left(\sum_{\bx,\by\in\supp Q^\prime}
I_A(\bx,\by)-I_B(\bx,\by)\right) \ \ .
\eqno(7.21)
$$
That the conditions of  Lemma 7.4  are satisfied
follows from the following observations:
An elementary change of a configuration $\om$ consists of the
addition or the removal of a single bond.  Such a change will either
join a pair of loops or cut a loop into two.  If $E_\om(Q)$ is
affected, the change is by not more than  $2S^{\vert\supp Q\vert}$.
However, the change is zero unless there is a pair of sites
$\bx,\by\in\supp Q$ whose loops are either joined or disconnected
in the process.
In this situation the value of the increasing function
$\sum_{\bx,\by\in\supp Q}
I_A(\bx,\by)-I_B(\bx,\by)$ is changed by at least 1.

Applying the Lemma we get:
$$\eqalignno{
&\left\vert\int \mu(\d\om) E_\om(Q)E_\om(Q^{\prime})
-\left(\int \mu(\d\om) E_\om(Q)\right)\left(
\int \mu(\d\om) E_\om(Q^{\prime})\right)\right\vert&\cr
&\quad\leq 4S^{\vert\supp Q\vert +\vert\supp Q^{\prime}\vert}
\sum_{C_1,C_2}\sum_{\bx,\by\in\supp Q\atop
\bu,\bv\in\supp Q^{\prime}}
\vert\langle I_{C_1}(\bx,\by) ; I_{C_2}(\bu,\bv)\rangle\vert &(7.22)\cr
}$$
which concludess the second step of the proof.

3)  Our goal now is to estimate the quantities
 $\langle I_{C_1}(\bx,\by) ; I_{C_2}(\bu,\bv)\rangle$ appearing
in the right side of (7.22).
By the
``$A$--$B$'' symmetry, it suffices to derive an estimate
in terms of the $A$-model's connectivity function $\tau_A(\cdot,\cdot)$.

The events $I_C(\bx,\by)$ are of two possible types:

i) two $A$-sites, $\bx_A$ and $\by_A$, are in the same connected
$A$-cluster, or

ii)  two $B$-sites, $\bx_B$ and $\by_B$, are {\it not}
separated by a connected $A$-cluster (\ie are connected by a
$B$-cluster).

In either case, the events are determinded by the $A$ cluster
of the set $[\bx_{C},\by_{C}]_A$ defined by:
$$\eqalignno{
[\bx_{A},\by_{A}]_A &=\hbox{the two point set }
\{\bx_{A},\by_{A}\}&(7.23)\cr
[\bx_{B},\by_{B}]_A &=\hbox{the collection of the $A$ sites in the interval
joining }\bx_{B} \hbox{ and } \by_{B}&(7.24)\cr
}$$

For the application of Lemma 7.5 we define two monotone increasing
events as follows: if $C_1=A$ put $I_1=I_A(\bx,\by)$ and $D_1
=[\bx_A,\by_A]_A$; if $C_1=B$ put $I_1=1-I_A(\bx,\by)$ and $D_1
=[\bx_B,\by_B]_A$. $I_2$ and $D_2$ are defined in terms of
$I_{C_2}(\bu,\bv)$ and $[\bx_{C_2},\by_{C_2}]_A$
in the same way. Lemma 7.5 then implies:
$$\eqalignno{
\vert\langle I_{C_1}(\bx,\by) ; I_{C_2}(\bu,\bv)\rangle\vert
&\leq\Prob{$[\bx_{C_1},\bu_{C_1}]_A\sim[\bu_{C_2},\bv_{C_2}]_A$}&\cr
&\leq\sum_{\bz\in [\bx_{C_1},\bu_{C_1}]_A\atop
\bz^\prime\in [\bu_{C_2},\bv_{C_2}]_A}
\Prob{$\bz_C\sim\bz^\prime_C$}&(7.25)\cr
}$$
where we have also used that $\vert\langle 1-I_1\,;I_2\rangle\vert
=\vert\langle I_1\,; I_2\rangle\vert$

Interchanging the r\^oles of $A$ and $B$ and using (7.3) one
finally obtains the estimate
$$
\vert\langle I_{C_1}(\bx,\by) ; I_{C_2}(\bu,\bv)\rangle\vert
\leq c(q)^{-1}
\min_{C=A,B}\sum_{\bz\in [\bx_{C_1},\bu_{C_1}]_C
\atop \bz^\prime\in [\bu_{C_2},\bv_{C_2}]_C}\tau_C(\bz,\bz^\prime)
\eqno(7.26)$$

Combining the inequalities (7.15), (7.18), (7.22), and (7.26),
one obtains  the estimate (7.5), for the case where $Q$ and $Q^\prime$
are products of $S^1$ and $S^3$ operators at distinct sites.

For the general case we use the identity
$S^2
=i[S^1,S^3]$ to express the product of spin operators
as a linear combination of products of only
$S^1$ and $S^3$.  The products may contain repeated factors.
With a trivial modification the argument given above
applies to such products as well.
\QEDD

\noindent
{\bf Proof of Theorem 7.1:}\nl
Theorem 7.1 is a direct consequence of Theorem 7.2.
\QEDD

It is interesting to consider the implications of the above analysis
for the spin chains with alternating coupling constants
$$
J_x=\cases{J_{\rm even}& if $x$ is even \cr
           J_{\rm odd}& if $x$ is odd \cr}
\eqno(7.27)$$
First it should be appreciated that while the spin system's Hamiltonian
is only periodic, the associated Potts models are translation invariant.
A great deal is known about such Potts models (though at  varying levels
of mathematical rigor), and our methods allow one to extract from that
some relevant information on the ground states of the quantum spin chains.

For the Potts models it is known that the correlation length is a meaningful
notion, defined by the limit
$$
\xi^{-1}_{\rm Potts}=-\lim_{x\to\infty}{1\over \vert x\vert}\ln \tau_C(0,x)
\eqno(7.28)$$
The existence of the limit is an easy consequence of the
supermultiplicativity (itself implied by an FKG type argument):
$\tau_0(x,z)\geq\tau_0(x,y)\tau_0(y,z)$,
where $\tau_0=c(q)^{-1}\tau_C$, $C=A$ or $B$.

Since Theorem 7.2 provides only upper bounds, let us first
strengthen the relationship between the long distance behavior of the
correlations in the two models.

\iproclaim Theorem 7.6.
For the models with alternating coupling constants (7.27):
if $\xi_{\rm Potts}<\infty$ then also
$$
\xi^{-1}_{\rm QSChain}=-\lim_{x\to\infty}{1\over \vert x\vert}\ln \vert\langle
S_0^3 S_x^3\rangle\vert
\eqno(7.29)$$
exists and $\xi_{\rm Potts}=\xi_{\rm QSChain}$.
\eproclaim
\proof:
Let $0<m=\xi^{-1}_{\rm Potts}$ be defined by the limit (7.28).
The theorem will follow from the following inequalities: for any
$\epsilon >0$ not too large, there exist constants $C_+,C_->0$ such that
$$
C_+ e^{-m(1-\epsilon)\vert x\vert}
\geq \vert\langle S^3_0, S^3_x\rangle\vert
\geq C_- e^{-m(1+\epsilon)\vert x\vert}
\eqno(7.30)$$
for all $x$ large enough. Since the spin-spin correlations are
dominated by the function $\tau(\cdot)$, the upper bound is trivial
(in fact, an auxilliary argument
shows that  $C(S) e^{-m |x|}$ would do).
We now provide the argument for the lower bound on the spin-spin
correlation function. For concreteness let us assume that there is
no percolation in the $A$ Potts model.

By (2.8) the problem amounts to estimating from below
$$
P(0,x)\equiv\Prob{$0$ and $x$ are on the same loop}
$$
We first show that for all small $\epsilon >0$ and finite $D>0$ ($D$
will be taken $\Order(1)$) the following quantity
satisfies an exponential lower bound as in (7.30)
$$
P_{\epsilon,D}(0,x)\equiv
\hbox{\rm Prob}\left({\hsize= 7.3cm\vbox to 1cm{\eightrm
\noindent there are sites $u\in [(-4\epsilon x)_A,0],
v\in[x,(x+4\epsilon x)_A]$, such that $u$ and $v$ are on the same loop and
each of the $A$-strips at the edges of the
two intervals does not contain any bonds at times $t\in[-D,D]$
}}\right)
\eqno(7.31)$$
with $x$  odd.
As the outer boundary of any $A$-cluster is a loop
for the spin chain we obviously have:
$$
P_{\epsilon,D}(0,x)\geq \hbox{\rm Prob}\left({\hsize= 8cm\vbox to 1cm{\eightrm
\noindent
there is an $A$-cluster intersecting both $[-4\epsilon x,0]$
and $[x, x+4\epsilon x]$
but not $[-4\epsilon x, x+4\epsilon x]^c$,
and each of the $A$-strips centered at $(-4\epsilon x)_A,
\tover12,x-\tover12$, and $(x+4\epsilon x)_A$ does not contain any
bonds at times $t\in[-D,D]$}}\right)
$$
and therefore
$$\eqalign{
P_{\epsilon,D}(0,x)\geq & \hbox{\rm Prob}\left({\hsize= 8cm\vbox to
.75cm{\eightrm
\noindent $\exists$ an $A$-cluster connecting
$[-\infty,0]$ and $[x,+\infty]$
and each of the $A$-strips centered at $(-4\epsilon x)_A,
\tover12,x-\tover12$, and $(x+4\epsilon x)_A$ does not contain any
bonds at times $t\in[-D,D]$}}\right)\cr
&\cr
&-\Prob{$\exists$ an $A$-cluster connecting
$[-\infty,0]$ and $[x+4\epsilon x,+\infty]$}\cr
&-\Prob{$\exists$ an $A$-cluster connecting
$[-\infty,-4\epsilon x]$ and $[x,+\infty]$}\cr
}\eqno(7.32)$$

i) By FKG and translation invariance
$$\eqalign{
& \hbox{\rm Prob}\left({\hsize= 8cm\vbox to .75cm{\eightrm
\noindent $\exists$ an $A$-cluster connecting
$[-\infty,0]$ and $[x,+\infty]$
and each of the $A$-strips centered at $(-4\epsilon x)_A,
\tover12,x-\tover12$, and $(x+4\epsilon x)_A$ does not contain any
bonds at times $t\in[-D,D]$}}\right)\cr
&\cr
&\qquad\geq \Prob{$\exists$ an $A$-cluster connecting
$[-\infty,0]$ and $[x,+\infty]$}\cr
&\qquad \times\Prob{the $A$-strip centered at $\tover12$
does not contain any bonds at times $t\in[-D,D]$}^4
}$$
and, using (7.28),
$$
\Prob{$\exists$ $A$-cluster connecting
$[-\infty,0]$ and $[x,+\infty]$}
\geq c(q)^{-1} \tau_A(0,x)
\geq C^\prime e^{-(1+\epsilon)m\vert x\vert}
\eqno(7.33)
$$

ii) The middle term in (7.32) can be replaced by the straightforward estimate
$$\eqalignno{
&\Prob{$\exists$ $A$-cluster intersecting both
$[-\infty,0]$ and $[x+4\epsilon x,+\infty]$}&\cr
&\quad\leq\sum_{y\leq 0, z\geq x}\tau_A(y,z)&\cr
&\quad\leq C^{\prime\prime}e^{-(1+2\epsilon)m\vert x\vert}&(7.34)\cr
}$$
where  $C^{\prime\prime}>0$ depends only on $m$, and it is assumed that
$0<\epsilon<\tover14$.
By symmetry, (7.34) applies also to the last term in (7.32).
Combining (7.32-33) we obtain the desired bound for $P_{\epsilon,D}(0,x)$: for
some $C>0$ (depending on $D$):
$$
P_{\epsilon,D}(0,x)\geq C(1-Ce^{-\epsilon m\vert x\vert})
e^{-(1+\epsilon)m\vert x\vert}
$$

In order to complete the proof we have to argue that the exponential lower
bound on $P_{\epsilon,D}(0,x)$ implies a similar lower bound on $P(0,x)$.
More precisely, we use the bound for $P_{\epsilon,D}(0,x)$, which we only
need for ``$0$'' even and $x$ odd, to obtain the desired estimate
on $P(y,z)$ with $y=-1$ or $y=0$ and $z=x$ or $z=x+1$. This covers all
possible combinations of parities for the two sites.

We use Lemma 7.7 given below to show that there is some $\alpha>0$
such that for $y=-1$ or $0$ and $z=x$ or $x+1$:
$$
\Prob{$y$ and $z$ are on the same loop
$\mid$ the event described in (7.31)}
\geq e^{-m\alpha\epsilon\vert x\vert}
\eqno(7.35)$$
from which the desired bound immediately follows.

The basis for the claim (7.35) are the following three observations:

i) The event described in (7.31) implies the existence of
a line (forming part of a loop) outside of the rectangular neighbourhoods
$B_1\equiv[(-4\epsilon x)_A,0]\times[-D,D]$ and
$B_2\equiv[x,(x+4\epsilon x)_A]\times [-D,D]$,
connecting a point of the form $(u,s)$ with a point $(v,t)$,
where $u\in [(-4\epsilon x)_A,0], v\in[x,(x+4\epsilon x)_A]$ and
$s,t=\pm D$.

ii) For every pair of points $(u,s)$ and $(v,t)$, on the boundary of
the boxes $B_1$ and $B_2$ respectively (see i)), there is a pair of
local events $\event_1$ and $\event_2$ in the boxes $B_1$ and $B_2$
such that under $\event_1$ $(u,s)$ is
connected by a line to $(-1,0)$ or $(0,0)$, ad libitum, and under $\event_2$
$(v,t)$ is connected with $(x,0)$ or $(x+1,0)$.
The events, whose choice depends on $u,s,v$, and $t$, are depicted in
Figure 3.

iii)  The conditional probability of the above mentioned local events
$\event_1$ and $\event_2$ occurring together -
conditioned on any explicit configuration in the complement of
 $B_1$ and $B_2$ - is not smaller than $e^{-m\alpha\epsilon\vert x\vert}$,
for some $\alpha > 0$.

More explicitely, the local events are constructed as follows (see Figure 3).
Assuming the line connecting $B_1$ and $B_2$ (in the complement
of $b_1$ and $B_2$), reaches $B_1$ at the point $\bu_0(u,D)$. The event
$\event_1(\bu_0)$ inside $B_1$ is then described by specifying that in the
vertical strips between $u$ and $0$ one sees alternatingly the following
picture: in the first, third, fifth, ... strip, counted starting
from $u$, there is no bond at times $0\leq t\leq D$ and at least one bond
at a time $-D\leq t \leq 0$; in the second, fourth, ... strip there is
at least one bond with $0\leq t\leq D$ and no bond with $-D\leq t \leq 0$.
The strip to the left of $u$ is required not to contain any bonds
at times $-D\leq t\leq D$.
It is then obvious that in the box $B_1$ there will be a line connecting
$(u,D)$ and $(-1,0)$ and $(0,0)$. The case $u=0$ is treated by a trivial
modification of the above prescription. The event $\event_2(\bv_0)$ in $B_2$
is defined in a similar way.

\vskip 7.5 truecm

\noindent
{\bf Figure 3:} A schematic representation of the local event $\event_1$ used
in the proof of Theorem 7.6. Crossed areas do not contain any bonds and
gray areas contain at least one bond.

Note that the random loop
measure $\mu$ conditioned on an arbitrary configuration outside any finite
domain in space-time is of the form stated in Lemma 7.7. The function
$f$ is the number of loops inside the finite volume taking into account
the connections in the configuration outside, and $q=2S+1$. It is then
obvious that $f$ always satisfies the bounded-gradient condition with
$a=b=1$ because the addition or removal of a single bond can change
the number of loops by at most one.

The indicator function of the event $\event_1\cap\event_2$ is of the form
$FG$ with $F$ increasing and $G$ decreasing.

It is useful to introduce the auxiliary events $\event_{\bu_0,\bv_0}$,
which form a partition of the event described in $(7.31)$: $\{\bu_0=(u,s),
\bv_0=(v,t)\}$ is the ``first'' pair of points on the top or bottom boundaries
of $B_1$ and $B_2$ connected by a line outside these boxes (``first'' \eg in
lexicographic order).
We also define $J_{\rm min(max)}=\min(\max)\{J_{\rm even}, J_{\rm odd}\}$
and put $D=1$.

The observations i-iii) and Lemma 7.7 then yield for $y=(-1,0)$ or $(0,0)$
and $z=(x,0)$ or $(x+1,0)$:
$$\eqalign{
&\Prob{$y$ and $z$ are on the same loop
$\mid$ the event described in (7.31)}\cr
&\quad\geq\sum_{{\bu_0\in\partial B_1\atop \bv_0\in\partial B_2}}
\Prob{$\event_1(\bu_0)\cap\event_2(\bv_0)\mid \event_{\bu_0,\bv_0}$}
\Prob{$\event_{\bu_0,\bv_0}\mid\ $
 the event described in (7.31)}\cr
&\quad\geq \left\{\int\rho^{J_{\rm min}/(2S+1)}(\d\om)\Ind{
there is at least one bond in the strip $[0,1]\times [0,1]$}\right\}^{2
[\epsilon\vert x\vert]}\cr
&\qquad\times\left\{\int\rho^{(2S+1)J_{\rm max}}(\d\om)\Ind{
there is no bond in the strip $[0,1]\times [0,1]$}\right\}^{2[\epsilon
\vert x\vert]}\cr
&\qquad\times\left\{\int\rho^{(2S+1)J_{\rm max}}(\d\om)\Ind{
there is no bond in the strip $[0,1]\times [-1,1]$}\right\}^2\cr
&\quad = e^{-4(2S+1)J_{\rm max}}\left\{(1-e^{-J_{\rm min}/(2S+1)})
e^{-(2S+1)J_{\rm max}}\right\}^{2
[\epsilon\vert x\vert]}\cr
&\quad\geq Ce^{-m\alpha\epsilon \vert x\vert}\cr
}$$
for some $\alpha>0$ independent of $\epsilon$ and $x$. We can now conclude
that for $\epsilon>0$ small enough one has
$$
P(0,x)\geq e^{-m\alpha\epsilon\vert x\vert} P_\epsilon(0,x)\geq
e^{-m(1+(\alpha+1)\epsilon)\vert x\vert}
$$
for all $x$ large enough, which is a lower bound of the form (7.30).
\QED

In the above argument we use the following Lemma, in which we refer to the
order structure on the space of configurations given by the inclusion relation:
$\om\geq \om^\prime$ iff the set of bonds in $\om$ contains the set of bonds
in $\om^\prime$.  The proof is a rather standard FKG-type argument, and is
therefore omitted here.

\iproclaim Lemma 7.7.
Let $\mu(\d\om)$ be a
probability measure of the form:
$$
\mu(\d\om)={\rho^J_\Lambda(\d\om) q^{f(\om)}\over
\int\rho^J_\Lambda(\d\om) q^{f(\om)}}
$$
where $\rho^J_\Lambda(\d\om)$ is a Poisson measure on configurations
$\omega$ in a finite volume $\Lambda$ (in our case, $\Lambda$ is a
subset of the of space-time), $q\geq 1$, and $f(\omega)$ is a function
of bounded gradient, in the sense that
there are constants $a,b\geq 0$ such that
$f+ a N$ is non-decreasing and $f-bN$ is non-increasing
with $N(\omega)$
the total number of time-indexed bonds in $\omega$.
Then, for any two functions, $F$ non-decreasing
and $G$ non-increasing,
which depend on $\omega$ in two disjoint subsets of
$\Lambda$ (one determining $F$ and the other determining $G$)
one has the following comparison inequalities, with expectation
values with respect to modified Poisson measures:
$$\eqalign{
\int\mu(\d\om)F(\om)G(\om)
&\leq\int\!\rho^{q^bJ}_\Lambda(\d\om)F(\om)
\int\!\rho^{q^{-a}J}_\Lambda(\d\om)G(\om)\cr
\int\mu(\d\om)F(\om)G(\om)
&\geq \int\!\rho^{q^{-a}J}_\Lambda(\d\om)F(\om)
\int\!\rho^{q^bJ}_\Lambda(\d\om)G(\om)
}\eqno(7.36)$$
\eproclaim

Assuming now the validity of all the results on Potts models  presented
in references \tref{\HKW,\KLMR,\KS,\Wu} (not all of which have been
derived rigorously), we obtain the following implications.

\item{i)} If $\Je\neq\Jo$, the ground state
is unique, with exponential deacy of correlations and a spectral gap.

A perturbative version of this statement, for
small (or large) enough ratio of the two couplings,
is contained in a theorem of Kennedy and Tasaki \tref{\KT}.

\item{ii)} The case  $\Je =\Jo  $ corresponds to self dual Potts
models, in which case there is a dichotomy, which we have discussed
above in Section 6.  Its manifestation
in the Potts
model language is: at the self dual point there either is a first
order phase transition with coexistence of the
ordered and disordered phases, or the transition is second order -
with a unique state at which the correlations decay by a power law.
The threshold value of $q$ is $q=4$, which corrsponds to $S=1/2$.
The implication of the Potts model calculation is:

\itemitem{ii.a}
For $S=1/2$, there is a unique state,
but with algebraic decay of correlations and no spectral
gap (equation (6.3) is satisfied).   For the spin system that was known
independently of the
Potts model result, by the Bethe Ansatz solution and a result of \tref\LSM .

\itemitem{ii.b}
For $S>1/2$ the translation
invariance is spontaneaously broken and there are two partially dimerized
ground states, translates of each other,
with exponential decay of correlations and a non-vanishing
spectral gap (the exact value of the spectral gap can be
calculated, see \tref{\Baxb,\Klu }).

\item{iii)}
In the case $S=\tover12$, for weakly staggered coupling constants,
$\Je = 1+\delta, \,\Jo  =1-\delta,\, 0<\delta << 1$,
the correlation length is finite but a divergent function of $\delta$.
The conjectured values  of the corresponding exponents of
the 4-state Potts model \tref\Wu, yield the following behaviour for
the ground state
energy per site $\epsilon$ and the correlation length $\xi$
$$
\epsilon-\epsilon_0\sim -\vert\delta\vert^{\tover43}\,,
\qquad \xi\sim\vert\delta\vert^{-\tover23}
\eqno(7.37)$$
(up to logarithmic corrections).

This behavior  of the spin $\tover12$ Heisenberg
antiferromagnetic chain with alternating coupling strengths was first
obtained by Cross and Fisher in their study of the spin-Peierls transition
\tref\CF . It has the implication that
when the ellastic deformations of the underlying
lattice are taken into account, the ground state of
spin-$\tover12$ Heisenberg antiferromagnetic chain
develops the spin-Peierls instability. (For a related rigorous
result see \eg \tref\KL ).

\vfill\eject

\beginsection Appendix {\rm I}. Quasi-state decomposition for quantum states

Often, basic properties of the state of a quantum system are
elucidated by presenting it as a convex combination of states with a
particularly simple structure.   In this work we find it useful to
consider a broader class of affine decompositions - into convex
combinations of what is called below quasi-states.  These are linear
functionals which are required to meet the positivity requirements
(which are part of the definition of a quantum state) only in their
restrictions to certain Abelian subalgebras.

Quasi-state decompositions made already an implicit
appearance in the  discussion of the itinerant ferromagnetism in
reference \tref\AL .
The utility of such decompositions there, stems from
the fact that the ferromagnetic (or antiferromagnetic) properties of a
system with rotation-invariant spin-spin couplings  can be expressed
through the correlation functions of a commuting family of the spin
variables (e.g., $\{\sigma^3_x\}$ with $x$ the site index and 3
referring to the third
component of the Pauli spin matrices). The structure of the restriction
of the state to such families is made particularly transparant by decomposing
it into a convex combination of states in which the correlations are
either ``0 or 1'', in the sense that to spins are either locked
in a parallel (or antiparallel) state, or are completely independent.
This is also the major characteristic of the Fortuin-Kasteleyn
representation of the Ising model \tref{\FKa,\For}.

The purpose of this section is to formulate this notion, and
discuss some of its general properties in the context of a simple
example.

\noindent
{\bf {\rm I}.a   Definition of quasi-states}\nl

\iproclaim  Definition I.1.
Let $\A$ be a sub-algebra of observables of a quantum
system.  A densely defined linear functional $\rho$ (providing the
expectation value for observables) is called a quasi-state relative to
$\A$, and we will say that it is well adapted to $\A$, if $\rho$ is:
\item{i)} normalized: $\rho(\idty)=1$\hfill {\rm (I.1)}
\item{ii)} positive on $\A$, \ie for each bounded $A \in \A$,
$\rho(A^*A)\geq 0$\hfill {\rm (I.2)}
\eproclaim

The property of states which is not required of quasi-states is
the general positivity, {\rm I}.e., the unrestricted validity of ({\rm I}.2).

For finite dimensional quantum systems, the observables form
matrix algebras, and the linear functionals take the form:
$$
\rho(A) = \Tr Q A
\eqno({\rm I}.3)$$
where $Q$ is a corresponding density matrix.   When $\A$  is the
maximal-Abelian algebra of matrices which are diagonal in a certain
common basis $\{\ket{\alpha_n}\}$,  the conditions on $Q$ under
which $\rho$ is a quasi-state relative to $\A$ are:  i) $\Tr Q=1$, and ii)
$Q$ has non-negative diagonal
elements,
$$
\langle\alpha_n\mid Q\mid\alpha_n\rangle\geq 0 \quad \hbox{for }
n=1,...
\eqno({\rm I}.4)$$
The last requirement  is considerably weaker than the
condition
ii$'$)  $Q \geq  0$  (as a matrix)
needed for $\rho$ to be a state.

\iproclaim  Remark I.2.
At this point it is natural to ask whether ({\rm I}.4) has any
$\A -$ independent content.  The answer is rather negative: any
hermitian matrix with $\Tr Q \geq 0$ satisfies ({\rm I}.4) in some basis.
\eproclaim

As we shall see, in certain situations a convex decomposition
of a state into quasi-states (relative to a naturally relevant algebra),
provides a great deal of insight into a state's structure.  Such
decompositions take the form:
$$
\rho = \sum_n p_n Q_n
\eqno({\rm I}.5)$$
with the $Q_n$ quasi states relative to a common algebra , and
$p_n$ weights satisfying:
$$
p_n\geq 0, \quad     \sum_n p_n = 1
\eqno({\rm I}.6)$$

\noindent
{\bf {\rm I}.b.   An elementary example}\nll

As the simplest demonstration of the notion introduced
above,  consider the system consisting of two spin-1/2 objects, with
the spin-spin ferromagnetic interaction:
$$
H = -\sig_1\cdot\sig_2
\eqno({\rm I}.7)$$
The ground state of the related antiferromagnetic Hamiltonian
($-H$) is given by the rank-one projection
$$
P_0=\ketbra{0}{0}
\eqno({\rm I}.8)$$
onto the singlet state ($\ket{0} = (\ket{\tover12,-\tover12}-
\ket{-\tover12,\tover12})/\sqrt{2}$),
where the total spin is
$S=0$. In this state one can safely say that $\sig_1=-\sig_2$, \eg ,
$$
\Tr P_0 (\sig_1+\sig_2)^2 = 0
\eqno({\rm I}.9)$$
The ground state of the ferromagnet is slightly less elementary.
It has the three-fold degeneracy of the space on which $S=1$.  The
corresponding projection is $P_1 =  \idty - P_0 = S$
(S takes here only the
values 0 and 1), and the (normalized) state operator is
$$
Q_+ =  \tover13  S
\eqno({\rm I}.10)$$
While the spins are as parallel as the uncertainty relations allow, it is
not true that  $\sig_1-\sig_2= 0$.  In fact,
$$
\Tr Q_+ (\sig_1-\sig_2)^2  =1 > 0
\eqno({\rm I}.11)$$
The fact that except for the ``zero-point fluctuations'' the spins are
basically aligned, is easily seen in the following quasi-state
decomposition of the ferromagnetic ground state:
$$
Q_+ = \tover23 (\tover14)\idty + \tover13
(\tover12 T)
\eqno({\rm I}.12)$$
where $T = 2 S -1 $ is the exchange operator  ($T\ket{a,b}=
\ket{b,a};\; a,b=\pm$).

Two basic observation here are:
\item{-}  For any non-zero vector $\en\in\Rl^3$ both $\tover14 \idty$
and  $\tover12 T$ are quasi-state
operators relative to the Abelian algebra $\A_\en$ generated by $\{
\sig_1\cdot\en,\sig_2\cdot\en\}$.
The former is actually a state operator.
\item{-} On any of the Abelian factors $\A_\en$, the quasi-states
corresponding to
$\tover14 \idty$
and to $\tover12 T$ represent two simple alternatives.
In the first state,
$\sig_1\cdot\en$ and $\sig_2\cdot\en$ form a pair of uncorrelated
variables (taking the values
$\pm\tover12$ independently of each other), whereas in the quasi-state
$\tover12 T$, $\sig_1\cdot\en$ and $\sig_2\cdot\en$ are locked together.

By the rotational symmetry, it suffices to derive these
statements  for $\en=(0,0,1)$, within the standard Pauli spin matrix
representation, which makes the first statement really obvious.  For
the second we note that the restriction of $\Tr(\tover12 T)A$  to
$A\in\A_{(0,0,1)}$ is
determined by only the diagonal elements of $\tover12 T$ in the basis
which diagonalizes $\{\sigma_1^3,\sigma_2^3\}$. That diagonal part of
$\tover12 T$ is
$\tover12(\ketbra{\tover12,\tover12}{\tover12,\tover12}\, +\,
\ketbra{-\tover12,-\tover12}{-\tover12,-\tover12})$.
In particular, it follows that the quasi-state  $\tover12 T$ yields the
following ferromagnetic analog of ({\rm I}.9):
$$
\Tr  \tover12 T  (\sig_1 -\sig_2)^2= 0
\eqno({\rm I}.13)$$

Thus, insofar as the restriction to  $\A_\en$ is concerned, the ground
state of the quantum ferromagnetic Hamiltonian is equivalent to a
positive temperature state of a classical ferromagnet, and the
decomposition ({\rm I}.12) is identical to the Fortuin-Kasteleyn random
cluster decomposition of the latter.

\noindent
{\bf {\rm I}.c.   Words of caution}\nll

Some notes of caution are due here.  The lack of full positivity
implies that quasi-states may lack some of the properties which are
familiar for the expectation value functionals $\langle\cdot\rangle$
associated with
regular states, such as:
\item{i)} $A \geq 0 \imply  \langle A\rangle\geq  0$\hfill ({\rm I}.14)
\item{ii)} $\langle A\rangle \leq \Vert A\Vert$   (the supremum norm of $A$)
\hfill({\rm I}.15)
\item{iii)} $\langle A*A\rangle = 0  \imply \langle AB\rangle  = 0$
for any (bounded) $B$ \hfill({\rm I}.16)

In fact, the quasi-state operator $\tover12 T$ provides us with the following
``counter-examples'':
\item{i)} $\Tr  \tover12 T  [(\sig_1-\sig_2)^2 -1] = -1 $,
while $(\sig_1-\sig_2)^2 -1\geq 0$\hfill ({\rm I}.17)
\item{ii)} $\Tr  \tover12 T  [2 - (\sig_1-\sig_2)^2] = 2$,
while $\Vert 2 - (\sig_1-\sig_2)^2\Vert =1$, \hfill ({\rm I}.18)

and
\item{iii)} $\Tr  \tover12 T (\sigma_1^3-\sigma_2^3)^2 = 0$, yet $\Tr
\tover12 T (\sigma_1^3-\sigma_2^3)\{T
\ketbra{-\tover12,\tover12}{-\tover12,\tover12}\} [=1]\neq 0 $.\hfill({\rm
I}.19)

These elementary assertions follow from ({\rm I}.13) and the observation
that $2 - (\sig_1-\sig_2)^2=2S-1=\pm 1$.

Nevertheless, the restriction of a quasi-state to any of the
algebras for which it is well adapted is free of the above ``pathologies''.
In particular, the three principles ({\rm I}.14)-({\rm I}.16) are satisfied as
long as $A$
and $B$ are restricted to any common $\A_\en$.

One may also note that while a quasi-state is not fully
characterized by its restriction to a single Abelian factor, in the above
case the operator $T$ is uniquely determined by its rotation invariance
and the values of $\Tr \tover12 T A$  for $A\in\A_\en$.

\noindent
{\bf Appendix II: Strictly monotone observables for Gibbs-fields
and the Rising Tide Lemma}

The FKG structure of the probability measures $\mu_{\rm even}(\d\om)$
and $\mu_{\rm odd}(\d\om)$ is quite essential for our analysis in Sections
4-7. Here we provide
some details concerning this structure. We also introduce the notion of weak
strict monotonicity and prove the Rising Tide Lemma which was needed in
Theorem 6.1 to show
that breaking of translation invariance ($\mu_{\rm even}
\neq \mu_{\rm odd}$) implies staggered behaviour of the nearest neighbor
spin-spin correlation ($\langle \es_x\cdot\es_{x+1}\rangle_{\rm even(odd)}
\neq \langle \es_{x+1}\cdot\es_{x+2}\rangle_{\rm even(odd)}$).

Recall from Section 3 the definition of the partial order on the set of
configurations
$\om=(\om_A,\om_B)$,  with $\om_{A (B)}$
the set of $A(B)$-bonds labeled
by space-time points.  The partial order is defined by declaring
$$
\om\preceq\om^\prime \qquad\hbox{ if } \om_A\supset\om^\prime_A \hbox{ and }
\om_B
\subset\om^\prime_B
\eqno({\rm II}.1)$$
That leads to the notions monotonicity of functions and
domination of measures: i) a function $f$ is called monotone increasing
 if
$$
f(\om)\leq f(\om^\prime) \qquad\hbox{ for all }
\om\preceq\om^\prime
\eqno({\rm II}.2)$$
ii) a probability measure $\mu(\d\om)$ is said to be {\it dominated\/}
by another probabiltiy measure $\nu(\d\om)$, denoted as $\mu(\d\om)
\preceq \nu(\d\om)$, if
$$
\int\mu(\d\om) \, f(\om)\leq\int \nu(\d\om) \,f(\om)
\qquad\hbox{ for all monotone increasing } f
\eqno({\rm II}.3)$$

This structure is useful in a number of ways.

1)  Fortuin, Kasteleyn and Ginibre \tref\FKG\  provide a general criterion,
under which a probability measure would have the property:
$$
\int\mu(\d\om) \, f(\om) g(\om) \geq\int\mu(\d\om) \, f(\om) \ \int\mu(\d\om)
\,g(\om)
\eqno({\rm II}.4)$$
for all monotone increasing  $f$  and $ g$.   The FKG condition is satisfied
by the random cluster measures (see \tref\ACCN ).

2)  For $\mu$ satisfying (II.4), the probability
measures conditioned on the configuration in the complement of
a finite volume, are the monotone in the sense:
$$
\E_\mu(\cdot\mid \om_{\Lambda^c})\leq\E_\mu(\cdot\mid \om^\prime_{\Lambda^c})
\ \hbox{ for all } \om_{\Lambda^c}\preceq \om^\prime_{\Lambda^c}
\eqno({\rm II}.5)$$

3)  There exist domination relations among the random cluster measures
corresponding to Potts models with various values of $q$ ($q$=1 corresponding
to independent percolation).  Applications of these  relations can be found
 in \tref\FK\ and \tref\ACCN.

4)   Of particular interest in this work is the domination relation
$$
\mu_{\rm even}(\d\om) \preceq \mu_{\rm odd}(\d\om)
\eqno({\rm II}.6)$$
which follows by a simple limiting argument from (II.5).

The last relation includes the statement
$\langle \es_0\cdot\es_1\rangle_{\rm odd}
\leq \langle \es_0\cdot\es_1\rangle_{\rm even}$.  Our
goal now is to explain why whenever the two states are  different,
the inequality holds in the strict sense.  A key role in the argument
is played by Lemma II.2, whose name draws on the observation
 that a rising tide lifts all the ships in the harbor.

It is useful  to introduce the notion of
{\it strict monotonicity\/}.

\iproclaim Definition II.1.
Let $\M$ be a set of probability measures on a configuration space
$\Omega$ with an order structure as above. A monotone increasing
function on $\Omega$ is {\rm strictly  increasing
 in the weak sense with respect to $\M$} (for short $\M$-strictly
increasing), if
$$
\int\mu(\d\om) \, f(\om)<\int \nu(\d\om) \,f(\om)
\qquad\hbox{ for all } \mu,\nu\in\M,\, \mu\preceq\nu \,\,
\hbox{ and } \mu\neq\nu \ .
\eqno({\rm II}.7)$$
\eproclaim

The difference between  the two random cluster measures
is due only to the boundary conditions (pushed to infinity).  They
share, however, a common rule for the finite volume conditional
distributions.  (A phenomenon exhibited
also by the family of the Gibbs equilibrium states at a first order
phase transition.)  In the terminology discussed in \tref\Geo\ the measures
have the same set of { \it specifications}.

\iproclaim Lemma {\rm II}.2 (The Rising Tide Lemma).
Let $\M$ be a family of probability measures with common
specifications   $\gamma =
 \{\E(\cdot\mid \om_{\Lambda^c})\}_\Lambda$.  A sufficient condition for
a function $f$ to be $\M$ - strictly monotone increasing is that beyond some
finite volume $\Lambda_0$ the conditional expectation is strictly
monotone increasing with respect to the boundary conditions:
$$
\E(f\mid\om_{\Lambda^c})<\E(f\mid\om^\prime_{\Lambda^c})
\eqno({\rm II}.8)$$
for all $\Lambda\supset\Lambda_0$ and all pairs of boundary conditions
such that
$$
\om_{\Lambda^c}\preceq\om^\prime_{\Lambda^c}
\hbox{\ \  and  \ \ \  }
\om_{\Lambda^c}\neq\om^\prime_{\Lambda^c} \ \mod \gamma \ ,
\eqno({\rm II}.9)$$
where the last inequality means that the induced conditional expectations
are not identical.
\eproclaim

\iproclaim  Remark II.3.
For a given set of specifications, the relevant notion of boundary
conditions consists of the equivalence classes of $\{\om_{\Lambda^c}\}$
defined by
$$
\om_{\Lambda^c}\cong \om^\prime_{\Lambda^c}\quad\hbox{ iff }\quad
\E(\cdot\mid \om_{\Lambda^c})=\E(\cdot\mid \om^\prime_{\Lambda^c})
\eqno({\rm II}.10) \  \  .$$
It is easy to see that  for the random cluster model the equivalence
class of  boundary configurations
is determined by specifying the connectivity relations of the boundary sites
via the connecting paths in $\Lambda^c$.  (Obviously many configurations
 in the complement of $\Lambda$
would be equivalent.)
\eproclaim

\noindent
{\bf Proof of the Lemma:}
Let $\mu$ and $\nu \in\M$,  $\mu \preceq\nu$.  Then, by the general
result of Holley (\tref\Hol ), there is a coupling
$\rho$  which is a probability
measure on the space of pairs $\{(\om,\om^\prime)\}$ with marginals
$\mu$ and $\nu$ which is supported on pairs with $\omega\preceq
\omega^\prime$.   Using first the conditional expectation formula
and then the coupling, we have:
$$\eqalignno{
\nu(f) - \mu(f) &=
\int\nu(\d\om_{\Lambda^c})\,\E(f\mid\omega_{\Lambda^c})-
\int\mu(\d\om^\prime_{\Lambda^c})\,\E(f\mid\omega^\prime_{\Lambda^c})&\cr
&=\int \rho(\d\omega_{\Lambda^c}\times \d\omega^\prime_{\Lambda^c})
\left(  \E(f\mid\omega_{\Lambda^c})-
\E(f\mid\omega^\prime_{\Lambda^c})\right)&({\rm II}.11)\cr
}$$
Assuming the measures are different, there is some finite volume
$\Lambda \supset\Lambda_0$ for which $\rho$
assigns  a positive measure to pairs of configurations which are inequivalent
 as boundary conditions
in $\Lambda^c$.    Since the integrand in (II.11) is strictly positive at such
points, and is non-negative in general, it follows that $\nu(f) - \mu(f) > 0$.
\QEDD

\iproclaim Lemma {\rm II}.3.
For the random cluster model, for any $\bx,\by\in\Ir\times\Rl$,
 the random variable $I_A(\bx,\by)$ satisfy the criterion of the
Rising Tide Lemma II.2.  (The set $\Lambda_0$
can be taken as any box with $\bx$ and $\by$ in its interior.)
\eproclaim
\proof:
Let
$\Lambda$ be a finite box containing $\bx$ and $\by$ in its interior, and let
$\om_{\Lambda^c}\preceq\om_{\Lambda^c}^\prime$.  Then
each of these configurations induces a partition of the  $A$- sites on the
boundary
 of
$\Lambda$ into clusters connected in the exterior.  The relation
between the configurations means that the boundary partition
corresponding to $\om_{\Lambda^c}$ is
a refinement of that corresponding to $\om_{\Lambda^c}^\prime$.
The difference in the conditional expectations of  $I_A(\bx,\by)$
 is the result of  two  effects:  i)  the induced measures in $\Lambda$
are different, and ii) the finite volume conditions under which
x and y are connected are different (since $\om^\prime$ provides
a better connected boundary).  Denoting by $\bI_{\om_{\Lambda^c}}$ and
 $\bI_{\om^\prime_{\Lambda^c}}$ the corresponding indicator functions
(of only $\om_\Lambda$),
we have:
$$\eqalignno{
&\E(I_A(\bx,\by)\mid\om^\prime_{\Lambda^c})
-\E(I_A(\bx,\by)\mid\om_{\Lambda^c})&\cr
&\quad =
\E(\bI_{\om^\prime_{\Lambda^c}}(\bx,\by)\mid\om^\prime_{\Lambda^c})
-\E(\bI_{\om_{\Lambda^c}}(\bx,\by)\mid\om_{\Lambda^c})&\cr
&\quad =
\E(\bI_{\om_{\Lambda^c}}(\bx,\by)\mid\om^\prime_{\Lambda^c})
-\E(\bI_{\om_{\Lambda^c}}(\bx,\by)\mid\om_{\Lambda^c})
+
\E(\bI_{\om^\prime_{\Lambda^c}}(\bx,\by)-\bI_{\om_{\Lambda^c}}
(\bx,\by)\mid\om^\prime_{\Lambda^c})&\cr
&\quad\geq
\E(\bI(\bx\sim\by \hbox{ w.r.t. } \om_{\Lambda^c}^\prime
\hbox{ and } \bx\not\sim\by\hbox{ w.r.t. } \om_{\Lambda^c})
\mid\om_{\Lambda^c}^\prime)&({\rm II}.12)\cr
}$$
When  the two configurations are not equivalent then, due to the
strict positivity of the measure, the last term is strictly positive.
\QED

In this paper the Rising Tide Lemma was applied in Section 6.
We expect it to be of  interst also in various other situations where the
FKG structure is of relevance.

\noindent
{\bf  Acknowledgements}

It is a pleasure to thank I. Affleck and E.H. Lieb for stimulating discussions
on  topics related to this work, and C.M. Newman and A. Gandolfi
for providing us with the reference \tref\GKR .

\noindent
{\bf References}

\vskip .5cm

\let\REF\doref

\REF ACCN \ACCN \Jref
       M. Aizenman, J.T. Chayes, L. Chayes, and C.M. Newman
       "Discontinuity of the Magnetization in One-Dimensional
        $1/|x-y|^2$ Ising and Potts Models"
        J. Stat. Phys. @50(1988) 1--40

\REF AKN  \AKN \Gref
        M. Aizenman, A. Klein, and C. Newman
        "Percolation methods for disordered quantum Ising models"
        preprint

\REF  AL \AL \Jref
        M. Aizenman and E.H. Lieb
        "Magnetic Properties of Some Itinerant-Electron
        Systems at $T>0$"
        Phys. Rev. Lett. @65(1990) 1470--1473

\REF AM \AM \Jref
        M. Aizenman and Ph.A. Martin
        "Structure of Gibbs States of One-Dimensional
        Coulomb Systems"
        Commun. Math. Phys. @78(1980) 99--116

\REF AN \AN \Gref
         M. Aizenman and B. Nachtergaele
         "Geometric aspects of quantum spin systems II"
         in preparation

\REF Aff \Aff \Jref
         I. Affleck
        "Exact results on the dimerization transition in SU(n)
         antiferromagnetic chains"
         J. Phys. C: Condens. Matter @2(1990) 405--415

\REF AffL \AffL     \Jref
     I. Affleck and E.H. Lieb
     "A proof of part of Haldane's conjecture on quantum spin chains"
     Lett. Math. Phys. @12(1986) 57--69

\REF Baxa \Baxa \Bref
        R.J. Baxter
        "Exactly Solvable Models in Statistical Mechanics"
        Academic Press, London 1982

\REF Baxb \Baxb \Jref
         R.J. Baxter
         "Magnetisation discontinuity of the two-dimensional Potts
          model"
          J.~Phys. A: Math. Gen. @15(1982) 3329--3340

\REF BBa \BBa \Jref
     M.N. Barber and M.T. Batchelor
     "Spectrum of the biquadratic spin-1 antiferromagnetic chain"
      Phys. Rev. @B40(1989) 4621--4626

\REF BBb \BBb \Jref
      M.T. Batchelor and M. Barber
      "Spin-s quantum chains and Temperley-Lieb algebras"
      J.~Phys. A: Math. Gen. @23(1990) L15--L21

\REF BK \BK \Jref
        R.M. Burton and M. Keane
        "Density and uniqueness in percolation"
         Commun. Math. Phys. @121(1989 ) 501--505

\REF CNPR \CNPR \Jref
        A. Coniglio, C.R. Nappi, F. Peruggi, and L. Russo
        "Percolation and Phase Transitions in the Ising Model"
        Commun. Math. Phys. @51(1976)315--323

\REF CS \CS \Jref
     J.G. Conlon and J-Ph. Solovej
      "Random Walk Representations of the Heisenberg Model"
      J. Stat. Phys. @64(1991) 251--270

\REF  CF  \CF \Jref
       M.C. Cross and D.S. Fisher
       "A new theory of the spin-Peierls transition with special
        relevance to the experiments on TTFCuBDT"
       Phys. Rev. @B19(1979)402--419

\REF DLS  \DLS  \Jref
        F.J Dyson, E.H. Lieb, and B. Simon
        "Phase transitions in quantum spin systems with isotropic
         and non-isotropic interactions"
         J. Stat. Phys. @18(1978)335--383

\REF FNW \FNW \Jref
        M. Fannes, B. Nachtergaele , and R.F. Werner
        "Finitely correlated states for quantum spin chains"
         Commun. Math. Phys. @144(1992) 443--490

\REF For \For \Gref
        C.M. Fortuin:
        "On the random cluster model II and III"
        {\it Physica\/} {\bf 58}, 393--418 (1972) and
        {\it Physica\/} {\bf 59}, 545--570 (1972)

\REF FK \FKa \Jref
       C.M. Fortuin and P.W. Kasteleyn
       "On the random cluster model I"
        Physica @57(1972) 536--564

\REF FK \FKG \Jref
       C.M. Fortuin,  P.W. Kasteleyn, and L. Ginibre
       "Correlation inequalities on some partially ordered sets"
       Commun. Math Phys. @22(1971) 89--103

\def\FK{\FKa,\FKG,\For}

\REF Fur  \Fur \Bref
       H. Furstenberg
       "Recurrence in Ergodic Theory and Combinatorial Number
        Theory"
        Princeton University Press, Princeton 1981

\REF  GKR  \GKR \Jref
        A. Gandolfi, M. Keane,  and L. Russo
        "On the uniqueness of the infinite occupied cluster in dependent
        two-dimensional site percolation"
        Ann. Prob. @16(1988)1147--1157

\REF Geo  \Geo \Bref
        H.-O. Georgii
        "Gibbs measures and phase transitions"
        W. de Gruytter, Berlin - New York 1988

\REF Gin \Gin \Jref
        J. Ginibre
       "Reduced density matrices for the anisotropic Heisenberg model"
        Commun. Math. Phys.  @10(1968)140--154

\REF  GA  \GA  \Jref
      S.M. Girvin and D.P. Arovas
       "Hidden topological order in integer quantum spin chains"
       Physica Scripta @T27(1989)156--159

\REF GM \GM \Jref
       D. Goderis and C. Maes
       "Constructing quantum dissipations and their reversible
        states from classical interacting spin systems"
       Ann. Inst. H. Poincar\'e @55(1991) 805--828

\REF  HK  \HK  \Jref
      M. Hagiwara and K. Katsumata
      "Observation of $S=\tover12$ degrees of freedom in an undoped $S=1$
linear
      chain Heisenberg antiferromagnet"
      J. Phys. Soc. Japan. @61(1992)1481--1484

\REF  HKal2   \HKalb  \Jref
      M. Hagiwara, K. Katsumata, I. Affleck, B.I. Halperin, and J.P. Renard
      "Hyperfine structure due to the $S=\tover12$ degrees of freedom in
       an $S=1$ linear chain antiferromagnet"
      J. Mag. Mag. Materials @104-107(1992)839--840

\REF  HKal1  \HKala  \Jref
      M. Hagiwara, K. Katsumata, J.P. Renard, I. Affleck, and B.I. Halperin
      "Observation of $S=\tover12$ degrees of freedom in an $S=1$ linear
      chain Heisenberg antiferromagnet"
      Phys. Rev. Lett. @65(1990)3181--3184

\def\NENP{\HKala,\HK,\HKalb}

\REF Hal  \Hal  \Jref
     F.D.M. Haldane
     "Continuum dynamics of the 1-D Heisenberg antiferromagnet:
      identification with the O(3) nonlinear sigma model"
     Phys.Lett. @93A(1983)464--468

\REF HKW \HKW \Jref
       A. Hintermann, H. Kunz, and F.Y. Wu
        "Exact Results for the Potts Model in Two Dimensions"
        J. Stat. Phys. @19(1978) 623--632

\REF Hol \Hol \Jref
        R. Holley
        "Remarks on the FKG Inequalities"
        Commun. Math. Phys. @36(1974)277--231

\REF Ken \Ken \Jref
       T. Kennedy
      "Long range order in the anisotropic quantum ferromagnetic Hei\-sen\-berg
       model"
       Commun. Math. Phys. @100(1985) 447--462

\REF KL \KL \Jref
      T. Kennedy and E.H. Lieb
      "Proof of the Peierls Instability in One Dimension"
      Phys. Rev. Lett. @59(1987)1309--1312

\REF KLS \KLS \Jref
      T. Kennedy, E.H. Lieb, and B.S. Shastri
      "Existence of N\'eel order in some spin 1/2 Heisenberg antiferromagnets"
      J. Stat. Phys. @53(1988)1019--1030

\REF KT \KT \Jref
      T. Kennedy and H. Tasaki
      "Hidden Symmetry Breaking and the Haldane Phase in S=1 Quantum
       Spin Chains"
       Commun. Math. Phys. @147(1992)431--484

\REF KiTh \KiTh \Jref
       J.R. Kirkwood and L.E. Thomas
       "Expansions and Phase Transitions for the Ground State of Quantum
        Ising Lattice Systems"
       Commun. Math. Phys. @88 (1983) 569--580

\REF {Kl\"ua} \Klua \Jref
          A. Kl\"umper
         "New Results for q-state Vertex Models and
          the Pure Biquadratic Spin-1 Hamiltonian"
          Europhys. Lett. @9(1989) 815--820

\REF {Kl\"ub} \Klub \Jref
          A. Kl\"umper
         "The spectra of q-state vertex models and related
          antiferromagnetic quantum spin chains"
          J.Phys. A: Math. Gen. @23(1990) 809--823

\def\Klu{\Klua,\Klub}

\REF  KLMR \KLMR \Jref
         R. Koteck\'y, L. Laanait, A. Messager, and J. Ruiz
         "The q-state Potts Model in the Standard Pirogov-Sinai Theory:
          Surface Tensions and Wilson Loops"
          J.~Stat.~Phys. @58(1990) 199--248

\REF  KS \KS \Jref
         R. Koteck\'y and S.B. Shlosman
         "First-Order Phase Transitions in Large Entropy Lattice Models"
          Commun. Math. Phys. @83(1982) 493--515

\REF LM \LM \Jref
        E.H. Lieb and D.C. Mattis
        "Ordering Energy Levels of Interacting Spin Systems"
        J.~Math.~Phys. @3(1962) 749--751

\REF LSM \LSM    \Jref
   E. Lieb, T. Schulz, and D. Mattis
   "Two soluble models of an antiferromagnetic chain"
   Ann.Phys.(NY) @16(1961)407--466

\REF Mat1 \Mata  \Jref
      T. Matsui
      "Uniqueness of the Translationally Invariant Ground State
      in Quantum Spin Systems"
      Commun. Math. Phys. @126(1990) 453--467

\REF Mat2 \Matb  \Jref
     T. Matsui
     "On Ground State Degeneracy of $\Ir_2$-Symmetric Quantum
      Spin Models"
      Pub. RIMS (Kyoto) @27(1991)657--679

\REF New \New \Jref
         C.M. Newman
          "A general central limit theorem for FKG systems"
         Commun. Math. Phys. @91(1983)75--80

\REF dNR \DNR \Jref
      M. den Nijs and K. Rommelse
      "Preroughening transitions in crystal surfaces and valence bond phases
      in quantum spin systems"
      Phys. Rev. @B40(1989) 4709--4734

\REF PS \PS \Jref
     L. Pitaevskii and S. Stringari
     "Uncertainty principle, quantum fluctuations and broken symmetries"
     J. Low Temp. Phys. @85(1991)377--388

\REF Sha  \Sha \Jref
      B.S. Shastri
     "Bounds for correlation functions of the Heisenberg antiferromagnet"
     J.~Phys. A: Math. Gen. @25(1992)L249--253

\REF Sut \Sut \Jref
     B. Sutherland
     "Model for a multicomponent quantum system"
     Phys. Rev. @B12(1975)3795--3805

\REF Suto \Suto \Gref
     A. S\"ut\"o
    "Percolation transition in the Bose gas"
     to appear in J. Phys. A: Math.~Gen.

\REF Tho \Tho \Jref
     L. E. Thomas
     "Quantum Heisenberg ferromagnets and stochastic exclusion
      processes"
     J. Math. Phys. @21(1980) 1921--1924

\REF Toth \Toth \Gref
        B. T\'oth
       "Improved lower bound on the thermodynamic pressure of the spin 1/2
        Heisenberg ferromagnet"
       to appear in Lett. Math. Phys.

\REF Wu \Wu \Jref
        F.Y. Wu
        "The Potts model"
        Rev. Mod. Phys. @54(1982) 235--268

\vfill\eject

\bye